\documentclass[twocolumn,nobibnotes,nofootinbib,superscriptaddress,preprintnumbers,aps,prd,10pt]{revtex4-1}
\pdfoutput=1

\usepackage{graphicx}
\usepackage{lpic}
\usepackage{latexsym}
\usepackage{amsfonts}
\usepackage{amssymb}
\usepackage{color}
\usepackage{amsmath}
\usepackage{slashed}
\usepackage{dcolumn}
\usepackage{feynmf}
\usepackage{verbatim}
\usepackage{multirow}

\definecolor{darkblue}{rgb}{0,0,0.5}
\usepackage[
pdfauthor={Wolfgang Altmannshofer, Patrick J. Fox,
Roni Harnik, Graham D. Kribs and Nirmal Raj}]{hyperref}

\DeclareGraphicsExtensions{.pdf,.png,.jpeg}

\definecolor{orange}{rgb}{1,0.5,0}

\newcommand{\weyli}{\chi_A}
\newcommand{\weylii}{\chi_B}
\newcommand{\modu}{Model~U}
\newcommand{\modd}{Model~D}
\newcommand{\modud}{Model~UD}

\newcommand{\gsim}{\gtrsim}
\newcommand{\lsim}{\lesssim}
\newcommand{\ra}{\rightarrow}
\newcommand{\ct}{c_\theta}

\newcommand{\sq}{\tilde{q}}
\newcommand{\sle}{\tilde{\ell}}
\newcommand{\Mchi}{M_\chi}
\newcommand{\delM}{\Delta{M}/M_\chi}		
\newcommand{\Mrat}{M_\phi/\Mchi}

\newcommand{\Msq}{M_{\tilde{q}}}
\newcommand{\Msu}{M_{\tilde{u}}}
\newcommand{\Msd}{M_{\tilde{d}}}
\newcommand{\Msl}{M_{\tilde{\ell}}}

\newcommand{\MLL}{m_{\ell\ell}}
\newcommand{\lsq}{\lambda_{\tilde{q}}}
\newcommand{\lsu}{\lambda_{\tilde{u}}}
\newcommand{\lsd}{\lambda_{\tilde{d}}}
\newcommand{\lsl}{\lambda_{\tilde{\ell}}}

\allowdisplaybreaks

\begin{document}	

\preprint{FERMILAB-PUB-14-485-T}

\title{Dark Matter Signals in Dilepton Production at Hadron Colliders}

\author{Wolfgang Altmannshofer}
\affiliation{Perimeter Institute for Theoretical Physics, Waterloo, ON, 
N2L 2Y5, Canada}

\author{Patrick J. Fox}
\affiliation{Theoretical Physics Department,
Fermilab, P.O. Box 500, Batavia, IL 60510, USA}

\author{Roni Harnik}
\affiliation{Theoretical Physics Department,
Fermilab, P.O. Box 500, Batavia, IL 60510, USA}

\author{Graham D. Kribs}
\affiliation{Department of Physics, University of Oregon, Eugene, OR 
97403, USA}

\author{Nirmal Raj}
\affiliation{Department of Physics, University of Oregon, Eugene, OR 
97403, USA}

\begin{abstract}

We show that new physics can show up in dileptonic events through its radiative contributions to the dilepton invariant mass, leading to unique ``monocline'' features in $m_{\ell\ell}$, as well as the angular distribution of the leptons.  We focus in particular on the case of dark matter with scalar messengers coupling it to the quarks and leptons.  Consistent thermal models require the dark matter to have masses of 100's GeV and have $\gsim 1$ couplings to the Standard Model (SM), implying that radiative corrections to the SM Drell-Yan rate can be sizeable.  We consider the case of Majorana, Dirac, and pseudo-Dirac dark matter and show that there are regions of parameter space where the non-existence of a monocline, which starts at roughly twice the dark matter mass, $m_{\ell\ell} \sim 2 m_\chi$, places the strongest constraint on the model. We make predictions for the sensitivities at the high luminosity 14 TeV LHC as well as a future 100 TeV proton-proton collider. We find that our dilepton signal is most sensitive when the mediator and the dark matter are nearly degenerate and conventional MET-based searches are least sensitive.
\end{abstract}

\maketitle

\section{Introduction}
\label{sec:intro}

Now that the Higgs has been discovered, one of the highest priorities for
the LHC in the next run is to find (or place strong bounds on)
particle dark matter. The standard approach is to look for
dark matter pair production as missing transverse momentum (MET)
in association with some initial state radiation.
Processes of that type could arise from effective operators~\cite{Cao:2009uw, Beltran:2010ww, 
Goodman:2010yf,Bai:2010hh,Goodman:2010ku, Fox:2011fx,
Rajaraman:2011wf,Fox:2011pm,Cheung:2012gi} 
or UV-complete simplified models involving various types
of mediators \cite{
Bai:2010hh, 
Fox:2011fx, 
Fox:2011pm, 
Goodman:2011jq, 
Shoemaker:2011vi,
Chang:2013oia,
An:2013xka,
Bai:2013iqa,
DiFranzo:2013vra,
Papucci:2014iwa,
Bai:2014osa,
Chang:2014tea,
Garny:2014waa,
Buchmueller:2014yoa,
Abdallah:2014hon}.
One of the principal results from these works is that there is
complementarity between the bounds from direct detection
and the bounds from the various types of LHC searches
for evidence of dark matter and its mediators.

The simplest models contain only a few parameters:
the dark matter mass, the mediator mass(es), and the
coupling(s) of the dark matter to one (or more) Standard Model (SM) field(s).
Consider the case where the dark matter
is a fermion, the mediator is a scalar (but with the $SU(3)\times SU(2)\times
U(1)$
quantum numbers of a Standard Model quark), and there is a
renormalizable interaction between a light quark, the dark fermion,
and the scalar mediator.
While there are several constraints, the dominant ones are~\cite{Chang:2013oia}:
\begin{itemize}
 \item[(i)] direct detection for small mass splittings between the dark fermion
and scalar mediator;
 \item[(ii)] jets $+$ MET constraints from LHC for large mass splittings caused by
scalar mediator production and decay to dark fermions and
jets.
\end{itemize}
These constraints tend to push the dark
fermion and scalar mediator masses to larger values with moderate
 mass splittings.
However, the dark matter annihilation cross section,
that sets the thermal relic abundance, scales with
positive powers of the coupling multiplying \emph{negative} powers
of the dark fermion mass (or scalar mediator mass -- it doesn't matter
since their mass scales are highly correlated).
The downward march of the experimental bounds must therefore be
accompanied by an upward march of the coupling constant(s),
which in some cases, can now be $\gsim 1$~\cite{Chang:2013oia}.

Couplings $\gsim 1$ provide a potential new avenue for
exploration and discovery at collider experiments.
Namely, they open up the possibility of experimentally
measurable radiative corrections of dark fermions and
mediators to Standard Model
processes.  There are several types of radiative corrections
that we could consider.  (Earlier work that has considered radiative
corrections of dark matter include~\cite{Dittmaier:2009cr, Freitas:2014jla,Harris:2014hga})
In this paper, we focus on the radiative corrections
to dilepton production at hadron colliders.
For the model, we assume there is a dark fermion
(that can acquire Dirac and Majorana masses) as well as
scalar messengers:
scalar quarks that couple to light quarks and the dark fermion with coupling strength $\lsq$,
and scalar leptons that couple to leptons and the dark fermion with coupling strength $\lsl$.

Dilepton production is well known to be a harbinger for new
physics (NP).  New gauge bosons ($Z'$s), extra dimensions, and
effective operators are well known examples that have already
been bounded by ATLAS~\cite{Aad:2014wca,Aad:2014cka} and CMS~\cite{Chatrchyan:2012oaa,CMS:2014aea} using the shape and normalization
of dilepton production as a function of the dilepton invariant mass,
$\sqrt{\hat{s}} = \MLL$.
Our primary interest is the new dark sector ``box'' contributions
to $q\bar{q} \rightarrow \ell^+\ell^-$, that are proportional to
$\lsq^2 \lsl^2/(16 \pi^2)$ in the amplitude, interfering
with the usual Drell-Yan contribution from the Standard Model.
New kinematical features in the dilepton invariant mass spectrum
arise at invariant masses of twice the dark matter mass,
$\sqrt{\hat{s}} \simeq 2 m_\chi$, from both the real part of the new physics box amplitude as well as
an imaginary part for $\sqrt{\hat{s}} > 2 m_\chi$.
Unlike a $Z'$ search, however,
the box
contribution does \emph{not} look anything like a resonance.
In fact, there can be both constructive and destructive
interference effects that depend on the model and the
strength of the couplings.  At large, but still perturbative
couplings (roughly $\lsq, \lsl \gsim 1.4$), we find that
the $|\mbox{box amplitude}|^2$ contribution dominates.  This leads to a
unique monocline\footnote{In geology, a step-like feature in rock strata
consisting of rapid rise and a gentle falloff.
A common example is the Waterpocket fold in Capitol Reef National Park, 
Utah, USA \cite{capitolreef}.}
feature in the dilepton invariant mass.  Standard ``bump-hunter''
approaches are not appropriate, and could miss an otherwise observable
feature in the spectrum.  Like a $Z'$ or extra dimension search,
nontrivial contributions to the forward backward asymmetry $A_\text{FB}$ are also present.
Unlike a $Z'$ or extra dimension search, there is further
nontrivial angular dependence that can potentially be uncovered using strategies
implemented in searches for the new physics contributions to the
dijet angular distribution
\cite{ATLAS:2012pu}.\footnote{We thank G. Perez for pointing this out to us.}

All of these features arise from the box function contribution
to the amplitude that, we stress, cannot be captured by
effective four-fermion operators.  Instead, it is crucial
to ``scan'' over finite $\sqrt{\hat{s}} = \MLL$
to uncover the dominant features of the box contribution
that appear for $\sqrt{\hat{s}} \gsim 2 m_\chi$.  Given that we expect
the mediator masses larger than but of the order of the dark matter mass
(to obtain the correct relic abundance with
non-perturbative couplings), there is no regime where the
dark matter or the mediator can be ``integrated out''
while leaving a finite signal.  Indeed, one of our most important
results is that the \emph{mass scale} of the dark fermions
appears as a kinematical feature in the radiatively corrected
dilepton invariant mass distribution.  This is a completely
distinct approach to measuring a putative dark matter particle
mass at a collider.

We say ``putative'' since we still have no collider
probe of the stability of the dark matter.
Indeed, we should emphasize that the signal we propose
to look for, namely kinematical features in the dilepton invariant
mass spectrum and angular distributions consistent with
radiative corrections from a new ``dark'' sector,
could arise from other new physics sectors that have
nothing to do with dark matter. 
In this work we focus on one concrete dark matter model.

We have
organized the paper as follows.
First, we present the model
in Sec.~\ref{sec:model}.  Next, we discuss the dark sector box contributions
to the dilepton invariant mass distribution in Sec.~\ref{sec:lhcsigns}, with angular
distributions discussed in Sec.~\ref{subsec:angdist}.
In Sec.~\ref{sec:constraints} we consider constraints on the model 
from collider searches, dark matter direct detection experiments, 
and the dark matter relic abundance. 
Then, we compare the sensitivity of the dilepton signal with these other constraints on the parameter space in Secs.~\ref{sec:constraintsummary} and~\ref{sec:future}.
Specifically, we find that
the 20 fb$^{-1}$ 8 TeV dataset from LHC experiments could constrain
a modest region of parameter space that, in some cases, is not
yet excluded by other constraints.  Once the LHC goes up to
14 TeV with larger luminosity, a much more substantial
region of the parameter space can be probed.
In addition to our projected sensitivities at 14 TeV,
we also briefly consider the impact of a 100 TeV collider,
finding that it has excellent sensitivity.

\section{The Model: Mixed (Pseudo-Dirac) Fermionic Dark Matter}
\label{sec:model}

\renewcommand{\arraystretch}{1.5}
\begin{table}
\begin{center}
\begin{tabular}{cccc}
\hline\hline
~~Field~~ &  ~~Spin~~  & ~~$SU(3)_c \otimes SU(2)_W \otimes U(1)_Y $~~ & ~~$ Z_2 $~~  \\
\hline
$ \chi_1, \chi_2 $ & $ \mathbf{1/2} $ & $\mathbf{(1, 1, 0)}$ & $\mathbf{-1}$\\
$ \tilde{u}  $ & $\mathbf{0}$ & $\mathbf{(3,1,\frac{2}{3})}$ & $\mathbf{-1}$  \\
$ \tilde{d} $ & $\mathbf{0}$ & $\mathbf{(3,1,-\frac{1}{3})}$ & $\mathbf{-1}$ \\
$ \tilde{\ell}=\tilde{e},\tilde{\mu} $ & $\mathbf{0}$ & $\mathbf{(1,1,-1)}$ & $\mathbf{-1}$ \\
\hline\hline
\end{tabular}
\end{center}
\caption{The field content of our model and the corresponding quantum numbers. To ensure the stability
of the dark matter candidate, the Lagrangian is assumed to be invariant under a $Z_2$ parity.}
\label{table:qn}
\end{table}

The model we propose consists of two SM singlet fermions $\chi_{1,2}$, as well as colored and uncolored scalars $\tilde u$, $\tilde d$ and $\tilde \ell$
for mediating the interactions between the singlet fermions and the SM fermions. The field content along with
their quantum numbers is summarized in Table~\ref{table:qn}.
We impose a $Z_2$ parity under which the dark matter fermions as well as the mediators are odd, while all SM fields are even.
In this way, the lighter SM singlet fermion is stable and therefore a dark matter candidate. We describe the singlet fermions with two two-component (Weyl) spinors $\weyli$ and $\weylii$. We allow for both Dirac and Majorana masses, a scenario that we refer to as ``mixed'' dark matter (recently discussed by two of us in a supersymmetric context in \cite{Kribs:2013eua}).  
In the case where the Majorana mass
is small compared to the Dirac mass, such a scenario is also referred to as pseudo-Dirac
dark matter \cite{Hsieh:2007wq,DeSimone:2010tf}.  
The Lagrangian is given in two-component language by
%
\begin{eqnarray}
 \mathcal{L} &=& 
       i \weyli^\dagger \bar{\sigma}^\mu\partial_\mu \weyli
       + i \weylii^\dagger \bar{\sigma}^\mu\partial_\mu \weylii 
       + \mathcal{L}_{\mathrm{DM} \, \mathrm{mass}} 
       \\
 & & \!\!\!\! - \sum_{q = u,d} |D_\mu \sq|^2 - \Msq^2 \sq \sq^* 
       - (\sqrt{2} \ \lsq \sq^* \weylii^\dagger q_R^\dagger + \text{h.c.}) 
       \nonumber \\
 & & \!\!\!\! - \sum_{\ell=e,\mu} |D_\mu \sle|^2 
       - \Msl^2 \sle \sle^* 
       - (\sqrt{2} \ \lsl \sle^* \weylii^\dagger \ell_R^\dagger + \text{h.c.})
       \, , \nonumber
\label{eq:simpLag}
\end{eqnarray}
%
%
where $q_R^\dagger$ and $\ell_R^\dagger$ are the 
right-handed components of the SM quarks and leptons
respectively that are $SU(2)_W$ singlets, $\tilde u$, $\tilde d$ and 
$\sle$ are the colored and uncolored scalar mediators and 
$M_{\tilde u}$, $M_{\tilde d}$ and $\Msl$ are their masses. 
In the Lagrangian we omitted quartic couplings involving the 
scalar mediators since they have negligible impact on the 
phenomenology we discuss below.

We make four assumptions about the model:
\begin{itemize}
\item[1.]  We assume $\chi_B$ interacts with the SM fermions through 
the mediators, while $\chi_A$ does not.  This type of interaction is 
loosely inspired by ``mixed'' gaugino supersymmetric 
models \cite{Kribs:2013eua}
where the gaugino interacts with the quarks and squarks,
while the fermionic Dirac partner does not.  Having said this,
we do not assume the interactions or masses are otherwise
supersymmetrizable.  This can be parameterized in the context of dimensionless supersymmetry breaking~\cite{Martin:1999hc}.
Taking the alternate route of allowing couplings 
for $\chi_A$ would tend to reshuffle the effective strength
of the couplings, and this does not change the qualitative results. 
The only exception to this is
the possibility of additional CP-violating phases in the couplings.
However, we do not consider any of the couplings within the model
to violate CP in this work, so this does not add anything to our discussion.
\item[2.]  We assume the hidden sector couples only to one or both of
$u_R$ and $d_R$.  It is crucial that we have couplings to the light
fermions, though the handedness and isospin is not particularly
important.  We could also generalize to couplings with all flavors 
of quarks and leptons, i.e. the mediator couplings $\lambda_{\tilde{u}_i}$,
$\lambda_{\tilde{d}_i}$ and $\lambda_{\tilde{\ell}_i}$ could be non-zero for 
all SM flavors~$i$.  This is strongly constrained by flavor changing 
neutral current processes.  For the purposes of this paper
we assume that the mediator couplings are aligned with the 
SM Yukawa couplings such that the colored mediators couple only to the 
first generation of right-handed quarks.\footnote{An alternative approach 
to control flavor changing 
neutral currents would be to introduce 3 generations of mediators. 
This would allow to implement a minimal flavor violation structure, 
such that the mediator couplings are diagonal in flavor space and 
each generation of mediators couples to only one generation of 
SM fermions. Yet another possibility which we do not explore would 
be to assume that dark matter carries 
flavor~\cite{Kile:2011mn,Batell:2011tc,Agrawal:2011ze,Agrawal:2014aoa,Kile:2014jea}.}  
We choose right-handed
quarks to allow us to separate the effects of a up-type mediator
from a down-type mediator.  Moreover, due to $SU(2)_W$ invariance, an exact
alignment would not be possible for couplings to the left-handed SM quark
doublets.
\item[3.] We assume the hidden sector couples only to right-handed electrons and muons, $e_R$ and $\mu_R$,
through their respective mediators $\tilde{e}$ and $\tilde{\mu}$
with no flavor-violating couplings.  This could be trivially
extended to include $\tau_R$, but since di-tau production
is considerably more difficult to measure accurately 
compared with di-electron or di-muon production, we only 
consider the latter.  
\item[4.] Finally, we assume there are no CP violating phases
in the mass and coupling parameters.  
\end{itemize}

The mass Lagrangian for the dark matter sector, 
$\mathcal{L}_{\mathrm{DM} \, \mathrm{mass}}$,
is given in two-component notation by
%
\begin{equation}
\mathcal{L}_{\mathrm{DM} \, \mathrm{mass}} \; = \;
\begin{pmatrix} \weyli & \weylii \end{pmatrix}
\begin{pmatrix} \Delta M & M_d \\ M_d & \Delta M' \end{pmatrix}
\begin{pmatrix} \weyli \\ \weylii \end{pmatrix}
+ \text{h.c.} ~,
\label{eq:pseudodiracmasslag}
\end{equation}
%
where $M_d$ is a Dirac mass and $\Delta M$ and $\Delta M'$ 
are Majorana masses. 
Although our fourth assumption above makes all mass terms real, 
we first, for completeness, present general results for the mass 
eigenstates.  From the mixing of $\weyli$ and $\weylii$,
the mass matrix above gets diagonalized by some unitary matrix $U$, 
and we obtain eigenmasses given by
\begin{eqnarray}
\nonumber \bar{M}^2_1 = \frac{1}{2} \bigg[ |\Delta M|^2 + |\Delta M'|^2
+ 2 |M_d|^2~~~~~~~~~~~~~~~~~~~~~~~~~~~~
\\
\nonumber - \sqrt{ 4 |\Delta M M_d^*+  \Delta {M'}^* M_d|^2 + 
(|\Delta M|^2 - |\Delta M'|^2)^2)} \bigg]
\\
\nonumber \bar{M}^2_2 = \frac{1}{2} \bigg[ |\Delta M|^2 + |\Delta M'|^2
+ 2 |M_d|^2~~~~~~~~~~~~~~~~~~~~~~~~~~~~
\\
\nonumber + \sqrt{ 4 |\Delta M M_d^*+  \Delta {M'}^* M_d|^2 + 
(|\Delta M|^2 - |\Delta M'|^2)^2)} \bigg]
\label{eq:eigmassessqd}
\end{eqnarray}

Given our assumption that the physical phase in the mass 
Lagrangian vanishes, the mass eigenstates are 
\begin{equation}
\begin{pmatrix} \chi_1 \\ \chi_2 \end{pmatrix} =
\begin{pmatrix} \cos \theta & \sin \theta \\
                            - \sin \theta & \cos \theta   	
\end{pmatrix}
\begin{pmatrix} \weyli \\ \weylii  \end{pmatrix} \, ,
\label{eq:eigstats}
\end{equation}
with mixing angle given by 
\begin{equation}
\cos \theta \; = \; \frac{1}{\sqrt{2}}
   \left(1 + \frac{\Delta M' - \Delta M}{\sqrt{(\Delta M' - \Delta M)^2
                                               + 4 M_d^2}}\right)^{1/2} \, .
\label{eq:MixAng}
\end{equation}
Given that only one dark fermion $\chi_B$ couples to the SM, 
we can further simplify these expressions.  
Specifically, we can take $\Delta M' = 0$, which implies the heavier
eigenstate $\chi_2 \simeq \chi_A$ is the one that decouples
from the SM\@.  This gives the correct Majorana limit, 
i.e., the lightest dark fermion is the one that maximally
couples to the SM\@. 
This was explored previously in the context of ``mixed gauginos'' 
in supersymmetry \cite{Kribs:2013eua}.
The mass eigenvalues simplify to
\begin{eqnarray}
\bar{M}_1^2 &=& M_d^2 + \frac{\Delta M^2}{2} 
- \Delta M \sqrt{M_d^2 + \frac{\Delta M^2}{4}} \nonumber \\
\bar{M}_2^2 &=& M_d^2 + \frac{\Delta M^2}{2} 
+ \Delta M \sqrt{M_d^2 + \frac{\Delta M^2}{4}} \, .
\label{eq:EigMass}
\end{eqnarray}
Note that in this limit $|\bar{M}_2| - |\bar{M}_1| = \Delta M$, and with our choice of
mixing matrix in Eq.~(\ref{eq:eigstats}) without any additional phases, 
$\bar{M}_1 < 0$ and $\bar{M}_2 > 0$. In order to avoid the frequent use of minus signs in the following,
we define $M_1 \equiv -\bar{M}_1,~M_2 \equiv \bar{M}_2$ such that $M_1, M_2 > 0$.  
By holding the lighter eigenmass $M_1$ constant,
we can interpolate between the Dirac and Majorana limits by
using $\Delta M$ as a control parameter. In particular, $\Delta M = 0$
gives us the pure Dirac limit with $\cos \theta = 1/\sqrt{2}$, and
$\Delta M \ra \infty$ corresponds to the pure Majorana limit
with $\cos \theta = 0$. 
We will see shortly that this method of
interpolation is most useful for studying the phenomenology of
pseudo-Dirac dark matter.

\subsection{Simplified Models}

In addition to the four assumptions about the structure of the model,
we will further simplify the parameter space in order to capture 
the main results of the paper.  We do this using 
``simplified models'', which take the model from the
previous section, and consider several distinct simplifying
assumptions about the parameters.  This is analogous to what is
regularly done by the LHC collaborations to examine the impact
of their experimental searches on, for example, low energy
supersymmetry.  

We consider three simplified models which are summarized in  
Table~\ref{tb:simpmods}.  
The difference among these models are:
\begin{itemize}
 \item \modu~has $\chi$ coupling exclusively to right-handed up quarks,
 \item \modd~has $\chi$ coupling exclusively to right-handed down quarks, and
 \item \modud~has $\chi$ coupling to both right-handed quarks 
of the first generation.
\end{itemize}
In all three models, the
colored and uncolored scalar mediators are taken degenerate with mass $M_\phi$,
and all fermion-scalar-dark matter couplings are assumed equal, denoted by $\lambda$.
The mass of the lighter dark fermion state is denoted by $\Mchi$ 
in all three simplified models.
The mass of the heavier dark fermion state is given by $\Mchi + \Delta M$.

\begin{table}
\begin{center}
\begin{tabular}{ccc}
\hline\hline
~~Model~~ &  Couplings  &  Mediator masses \\
\hline
 U &  $\lambda \equiv \lsl = \lsu$~, &$ M_\phi \equiv \Msl = \Msu $ \\

   & $ \lsd = 0 $ & \\
\hline
 D & $ \lambda \equiv \lsl = \lsd$~, &$ M_\phi \equiv \Msl = \Msd $ \\
   & $ \lsu = 0 $ & \\
\hline
 UD & ~~$ \lambda \equiv \lsl = \lsu = \lsd $~~ & ~~$ M_\phi \equiv \Msl = \Msu = \Msd$~~ \\
   \hline\hline
\end{tabular}
\end{center}
\caption{The simplified models considered in the paper.}
\label{tb:simpmods}
\end{table}

\section{Dilepton Signatures}
\label{sec:lhcsigns}

\subsection{Overview}
\label{subsec:thresholdFX}

\begin{figure*}
\begin{center}
\includegraphics[width=17cm]{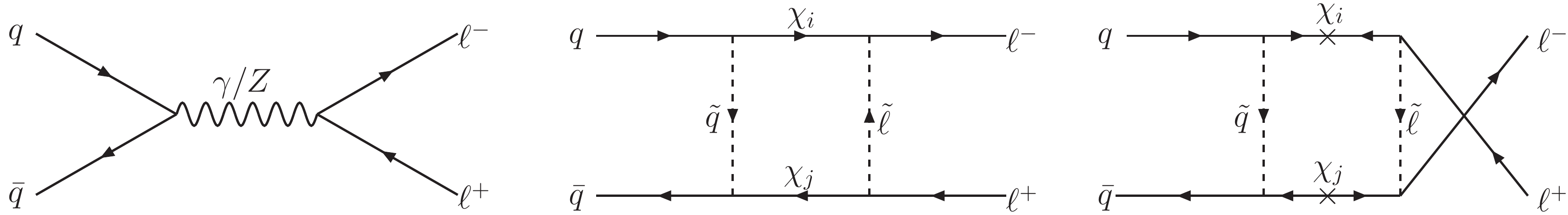}
\caption{Feynman diagrams of the most important processes that contribute to dilepton production in our model. The tree-level
$s$-channel photon-mediated and Z-mediated diagrams in the SM (left)
interfere with the standard box diagrams (center) and the crossed box diagrams (right).
The indices on the dark fermions are $i=1,2$ and $j=1,2$, thus making four combinations each of
standard and crossed box diagram. 
}
\label{fig:AllTheFeynmen}
\end{center}
\end{figure*}

At the LHC, dilepton production, $pp \to \ell^+\ell^-$, 
is dominated by the Drell-Yan process, $q\bar{q} \to \ell^+\ell^-$, 
with subdominant contributions from the production of tops, 
dibosons, dijets and W+jet.  Since our interest is in new physics
contributions that interfere with Drell-Yan, we neglect
these subdominant processes when computing Standard Model rates. 
This is a good approximation for at least LHC energies.  
We also do not incorporate QCD or electroweak NLO corrections,
since consistency would require also incorporating these
corrections to the new physics contribution, and this is beyond
the scope of this paper.  Hence, Standard Model dilepton production
is approximated solely by the tree-level $s-$channel photon-- and 
$Z$--mediated contributions shown in the left diagram 
of Fig.~\ref{fig:AllTheFeynmen}.  (At least some of the NLO corrections
would be common to both Drell-Yan and our new physics contribution, 
dropping out of the ratio.)   We also evaluate the couplings at a fixed
scale in perturbation theory.  RG improvement is straightforward
to incorporate, but does not significantly affect our results
other than redefining the new physics couplings 
$\lambda_{\tilde{q}},\lambda_{\tilde{\ell}}$ relative to
the modest RG evolution of the electroweak couplings. 

 At the one-loop level of our model, the dark fermions and mediators give corrections to dilepton production
$pp \to \ell^+\ell^-$ through self-energy corrections, vertex corrections and box diagrams.
The box diagram is enhanced relative to the self energies and the vertex corrections by a factor $\lambda^2/g^2$, where $g$ is an electro-weak coupling. As we will see below, in the interesting regions of parameter space that can be probed at current and future hadron colliders, the coupling $\lambda$ is considerably larger than the electroweak couplings $g$, $g'$.  Self-energy and vertex correction amplitudes can be safely neglected.
   The gauge boson self-energy diagrams would contribute to the running of the
 electroweak coupling at scales above the masses of the dark states. Ref. \cite{Alves:2014cda}
 discusses methods to probe hidden sectors at high energy scales by measuring deviations of
 the electroweak running from the SM at lower energies. Our approach, by taking only the
  box diagrams into account, probes the new physics sector directly at the mass scales of the particles involved. This is done by means of examining threshold effects, i.e., new terms
  in the amplitude that appear when states running in
  the loop go on-shell. We briefly review some salient aspects of these effects here. For a
  comprehensive review of dispersion relations in Feynman amplitudes, see \cite{Remiddi:1981hn}.

  Consider a general one-particle irreducible
   one-loop diagram of a $2\ra 2$
  scattering process. Let the masses of the propagator states
  that connect the initial and final states be $M_n$. 
  The amplitude develops an imaginary part for $\sqrt{s}>\sum_n |M_n|$,
  where $s$ is the Mandelstam variable. This imaginary part is given
  by the optical theorem, which states that
\begin{eqnarray}
\nonumber && 2\, \text{Im} \mathcal{M}(\mathbf{in}\ra \mathbf{out}) = \\
&& \sum_n \int d \Pi_n \mathcal{M}^*(\mathbf{out}\ra n) \mathcal{M}(\mathbf{in} \ra n)\,, 
\label{eq:genopticalthm}
\end{eqnarray}
 where \textbf{in} and \textbf{out} are the initial and final states respectively,
 $n$ denotes the intermediate on-shell states and $\int d\Pi_n$ is the
 integral over the phase space of $n$.

 When applied to the box diagram shown in the center of Fig.~\ref{fig:AllTheFeynmen}, 
 the imaginary part appears in the amplitude for 
$\sqrt{s}> M_{\chi_i} + M_{\chi_j}$
 and Eq.~(\ref{eq:genopticalthm}) becomes
\begin{eqnarray}
 && 2\, \text{Im}\mathcal{M}(q\bar{q}\ra \ell^+\ell^-) = \nonumber \\
&&  \sum_\chi  \int d \Pi_{\chi\chi} \mathcal{M}^*(\ell^+\ell^-\ra \chi\chi) \mathcal{M}(q \bar q\ra \chi\chi)\,.
\label{eq:boxopticalthm}
\end{eqnarray}
%
In addition to the turn-on of Im$\mathcal{M}$,\footnote{Note 
that even if we allowed the masses $\Delta M, \Delta M', M_d$ and couplings $\lsq$, $\lsl$ in Eq.~(\ref{eq:pseudodiracmasslag})
to be complex, no extra phase would appear in $\mathcal{M}_{\text{box}}$, as only absolute values of these quantities enter: $\mathcal{M}_{\text{box}} \propto |\lsq|^2|\lsl|^2$.}
the real part of the amplitude, Re$\mathcal{M}$, undergoes a
  continuous but sharp rise as well, a consequence of the dispersion relations that follow from the
  unitarity of the $S$-matrix  \cite{Remiddi:1981hn}.

Since the couplings of our model are only to right-handed SM fermions, the new physics amplitude interferes only with that part of the SM amplitude involving
right-handed external fermions. That is, if we denote the Standard Model amplitude by,
\begin{equation}
\mathcal{M}_{\text{SM}} = \mathcal{M}_{\text{SM}}^{{\text{LL}}} + \mathcal{M}_{\text{SM}}^{\text{LR}} + \mathcal{M}_{\text{SM}}^{\text{RL}} + \mathcal{M}_{\text{SM}}^{\text{RR}} ~,
\label{eq:chiralamps}
\end{equation}
where the first (second) letter of each superscript denotes the chirality of the initial state quark (final state lepton), then only $\mathcal{M}_{\text{SM}}^{\text{RR}}$ interferes with the new physics 
contributions given our assumptions about how the new fermions
couple in the model. Including the corresponding ``left-handed'' mediators would allow interference with all of the terms above.

\subsection{Dilepton Rates: Dirac Case}
\label{subsec:diracrates}

We now discuss the role of interferences and threshold effects in
generating the various signatures of our model. 
We first consider the simple case of a dark matter
candidate that is a Dirac fermion.

The only box diagram that contributes in this case is shown in the center of
Fig.~\ref{fig:AllTheFeynmen}. 
We can then write the total amplitude at the parton level as
\begin{equation}
\mathcal{M}_{\rm{total}} = \mathcal{M}_{\rm{SM}} +\mathcal{M}_{\rm{box}} \, ,
\end{equation}
where the Standard Model amplitude $\mathcal{M}_{\text{SM}}$
corresponds to the sum of the $s$-channel photon- and $Z$-mediated 
tree-level amplitudes with all polarizations shown in the 
left diagram of Fig.~\ref{fig:AllTheFeynmen},
\begin{equation}
\mathcal{M}_{\rm{SM}} = \mathcal{M}_{\rm{photon}} + \mathcal{M}_{\rm{Z}} \, .
\end{equation}
%
Neglecting the masses of the quarks and leptons, we can write the
double differential parton level $q\bar q \to \ell^+\ell^-$ cross section as
  \begin{eqnarray}
  d\sigma_{\text{total}} &\equiv& \frac{d^2\sigma_{\text{total}}}{d\cos\theta d\MLL}  
  \label{eq:XStotalDirac} \nonumber 
  \\[4pt] &  =& d\sigma_{\text{SM}} + d\sigma_{\text{int}} + d\sigma^{\text{Re}}_{\text{box}} + d\sigma^{\text{Im}}_{\text{box}}~. \label{eq:sigma_tot}
  \end{eqnarray}
     Here, $\theta$ is the angle between the outgoing dilepton axis
     and incoming diquark axis in the center-of momentum frame.
     The terms in Eq.~(\ref{eq:sigma_tot}) are given by
\begin{eqnarray}
d\sigma_{\text{SM}} &=&  \frac{1}{32\pi s}|\mathcal{M}_{\text{SM}}|^2  ~, 
  \label{eq:sigmadefsdirac1} \\
d\sigma_{\text{int}} &=&  \frac{1}{32\pi s}
  2\text{Re}(\mathcal{M}_{\text{SM}}^{\text{RR}}\mathcal{M}_{\text{box}}^*) ,
  \label{eq:sigmadefsdirac2} \\
d\sigma^{\text{Re}}_{\text{box}} &=& \frac{1}{32\pi s}
  |\text{Re}\mathcal{M}_{\text{box}} |^2  ~,  
  \label{eq:sigmadefsdirac3} \\
d\sigma^{\text{Im}}_{\text{box}} &=& \frac{1}{32\pi s}
  |\text{Im}\mathcal{M}_{\text{box}} |^2 ~,
  \label{eq:sigmadefsdirac4}
\end{eqnarray}
%
where $\mathcal{M}_{\text{SM}}^{\text{RR}}$ is defined in Eq.~(\ref{eq:chiralamps}).
Our analytic results for the box contributions to the parton level cross section are collected in Appendix~\ref{app:xsections}.

\begin{figure}
\begin{center}
\includegraphics[width=8cm]{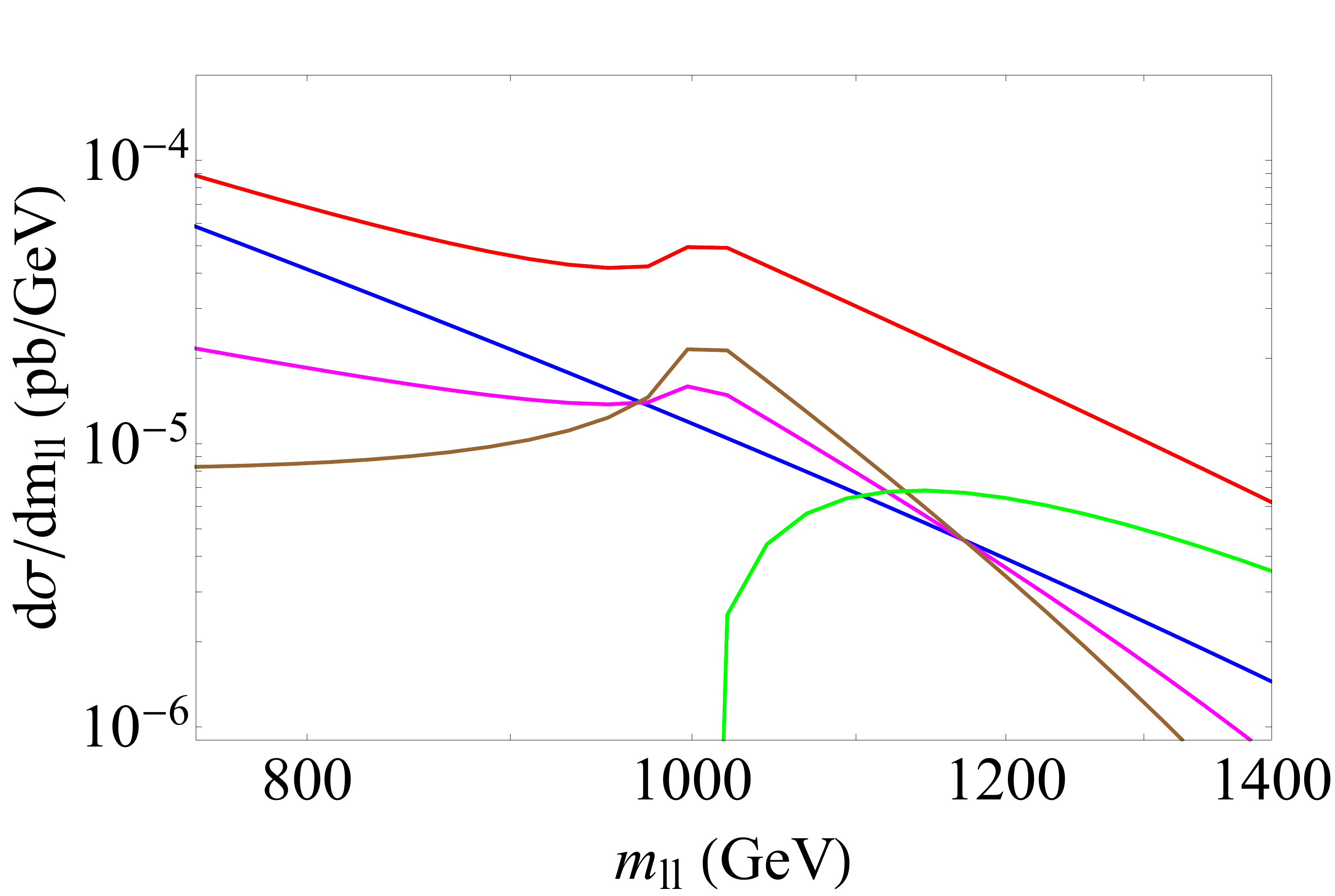}
\caption{The differential $pp \to \ell^+\ell^-$ cross sections
as a function of the dilepton invariant mass in \modu~with $\Delta M = 0$ (pure Dirac
limit), $\lambda = 1.8$ and $M_\chi=M_\phi=500$ GeV.
Here, blue: $d\sigma_{\text{SM}}$, brown:
 $d\sigma^{\text{Re}}_{\text{box}}$, green: $d\sigma^{\text{Im}}_{\text{box}}$,
 magenta: $d\sigma_{\text{int}}$,
 red: $d\sigma_{\text{total}}$, where these quantities are defined in
 Eqs.~(\ref{eq:XStotalDirac}) and~(\ref{eq:sigmadefsdirac1})~-~(\ref{eq:sigmadefsdirac4}).
}
\label{fig:DiracXS}
\end{center}
\end{figure}

    As we vary the dilepton invariant mass $\MLL$, we expect, for $\MLL \ll 2M_\chi$, $d\sigma_{\text{total}}$ to mimic
    the behavior of a non-resonant process generated by a higher-dimensional contact operator.
       The effects of such contact operators in dilepton production are being searched for by CMS~\cite{CMS:2014aea} and ATLAS~\cite{Aad:2014wca}. As we approach the kinematic threshold, $\MLL = 2M_\chi$, the contact operator description breaks down and a ``monocline'' feature arises from the contributions to: 
\begin{itemize}
 \item[(i)] $d\sigma_{\text{int}}$, due to threshold effects in $\text{Re}\mathcal{M}_{\text{box}}$,
 \item[(ii)]$d\sigma^{\text{Re}}_{\text{box}}$, which for sizeable couplings $\lambda$ can dominate over $d\sigma_{\text{int}}$ due to its containing eight
    powers of the coupling against four, and
 \item[(iii)] $ d\sigma^{\text{Im}}_{\text{box}}$, which turns on at $ \MLL \geq 2M_\chi$.
\end{itemize}
 We illustrate this behavior with an example in Fig.~\ref{fig:DiracXS},
 which shows the differential $pp \to \ell^+\ell^-$ cross section integrated over $\cos\theta$
 for \modu~at the LHC with 8~TeV center of mass energy. To obtain the proton level $pp \to \ell^+\ell^-$ cross section, throughout this work, we convolute the parton level results from Appendix~\ref{app:xsections} with MSTW2008NNLO parton distribution functions~\cite{Martin:2009iq}.
In the plot we set the mass splitting of the dark fermions to $\Delta M = 0$, corresponding
to the pure Dirac limit. The mediator couplings are set to $\lambda=1.8$ and we chose the masses of the dark fermions and the mediators as $M_\chi=M_\phi=500$~GeV\@.
The various curves correspond to 
  $\sigma_{\text{SM}}$              (blue);
  $\sigma^{\text{Re}}_{\text{box}}$ (brown);
  $\sigma^{\text{Im}}_{\text{box}}$ (green); 
  $\sigma_{\text{int}}$             (magenta);
  $\sigma_{\text{total}}$           (red); 
where these quantities are defined in Eqs.~(\ref{eq:XStotalDirac}) and~(\ref{eq:sigmadefsdirac1})~-~(\ref{eq:sigmadefsdirac4}).

Note that the example point shown in Fig.~\ref{fig:DiracXS} falls in a region of parameter space where
the new physics signal is dominated by $d\sigma_{\text{box}} \propto |\mathcal{M}_{\text{box}}|^2$.
At lower couplings, the dominant contribution to the signal becomes the interference term $d\sigma_{\text{int}}$
as defined in Eqs.~(\ref{eq:sigmadefsdirac2}). Numerically, we find that these
two regimes are separated by $\lambda \simeq 1.4$ in the presence of a
pure Dirac fermion.  
This comes into consideration when
we deal with constraints on our model from dilepton spectrum measurements and in
projecting results for future colliders.

 The blue, magenta and brown curves in Fig.~\ref{fig:DiracXS}
   (corresponding to $d\sigma_{\text{SM}}$, $d\sigma_{\text{int}}$, 
and $d\sigma^{\text{Re}}_{\text{box}}$ respectively) appear to 
intersect at $\MLL \sim 950$ GeV and $\MLL \sim 1150$ GeV. 
This intersection is a coincidence for the parameters presented and not a physical effect of 
our model. It arises
from the difference in which initial states contribute to $\mathcal{M}_{\text{SM}}$ and $\mathcal{M}_{\text{box}}$.  
Since both up and down quarks contribute to $\mathcal{M}_{\text{SM}}$, both these PDFs
 are convolved with the partonic level rates to obtain $d\sigma_{\text{SM}}$. In Model U, only the up quark contributes to $\mathcal{M}_{\text{box}}$, hence its PDF alone is
convolved with the partonic rates to obtain $d\sigma_{\text{int}}$ and
$d\sigma^{\text{Re}}_{\text{box}}$. Therefore, the apparent intersection seen here would 
be absent if we had presented partonic level rates, or used Model D or Model UD for illustration in
 Fig.~\ref{fig:DiracXS}.  Furthermore, with model U if the coupling is increased (decreased) the point where magenta and brown curves intersect moves up (down), and will not lie on the SM curve.  Similarly, if $M_\phi$ is altered the triple intersection would go away.

\subsection{Dilepton Rates: Mixed (Pseudo-Dirac) Case}
\label{subsec:pDiracRates}

Since a mixed dark matter candidate can be written as two 
Majorana eigenstates, we first begin with a brief discussion 
of the Majorana limit, that will be useful in understanding the
pseudo-Dirac case.  In addition to the standard box diagram, 
Majorana fermions have a ``crossed box'' diagram 
(with clashing fermion flow arrows) contributing at 
one-loop order, as shown by the right diagram in 
Fig.~\ref{fig:AllTheFeynmen}. 
The total amplitude becomes the sum
\begin{equation}
\mathcal{M}_{\text{total}} = \mathcal{M}_{\text{SM}} + \mathcal{M}_{\text{box}} + \mathcal{M}_{\text{xbox}} ~,
\label{eq:majamp}
\end{equation}
   where $\mathcal{M}_{\text{xbox}}$ is the amplitude for the
   crossed box diagram. Importantly, $\mathcal{M}_{\text{xbox}}$ comes with a minus
   sign relative to $\mathcal{M}_{\text{box}}$ due to the different ordering of the
   external spinors. Thus, the direct and crossed box diagrams
   interfere destructively, and we expect the new physics effects in the
   cross section to be much less pronounced in the Majorana case than in
   the Dirac case. In particular, we find that over large parts of the parameter space the ``monocline'' feature noticed in the Dirac scenario is washed out by the destructive interference.
   Even for sizeable couplings $\lambda \gtrsim 1.4$, the largest contribution to the deviation from the Standard
    Model cross section comes typically from the interference term between the tree and box amplitudes,
    which carries only four powers of the coupling~$\lambda$.

We now turn to the most general case of mixed (pseudo-Dirac) dark matter.
Four contributions arise from direct box diagrams
and four additional contributions from the crossed box diagrams, 
corresponding to the four combinations of $\chi_1$ and $\chi_2$ 
in the loop, as shown in Fig.~\ref{fig:AllTheFeynmen}.
The total amplitude is now given by
\begin{equation}
\mathcal{M}_{\text{total}} = \mathcal{M}_{\text{SM}} + \sum_{i=1,2} \sum_{j=1,2} (\mathcal{M}^{ij}_{\text{box}} + \mathcal{M}^{ij}_{\text{xbox}})~,
\label{eq:pDiracamp}
\end{equation}
where $\mathcal{M}^{ij}_{\text{(x)box}}$ is the (crossed) box amplitude with $\chi_i$
in the upper fermion propagator and $\chi_j$ in the lower fermion propagator. It is
illustrative to inspect the analytical form of the direct and crossed box amplitudes:
%
\begin{subequations}
\begin{eqnarray}
 \nonumber \mathcal{M}^{ij}_{\text{box}} &\propto&
 [\bar{u}(p_4)\gamma_\mu P_R u(p_1)][\bar{v}(p_2) \gamma_\nu P_R v(p_3)] \nonumber \\ && \times \int \frac{d^4q}{(2\pi)^4} \frac{q^\mu (q + p_1 + p_2)^\nu}{D_{ij}} ~,
\label{eq:Mijbox} \\[8pt]
 \mathcal{M}^{ij}_{\text{xbox}} &\propto& [\bar{u}(p_3) P_R u(p_1)] [\bar{v}(p_2) P_L v(p_4)] \nonumber \\
&& \times \int \frac{d^4q}{(2\pi)^4} \frac{\bar{M}_i \bar{M}_j}{D_{ij}} ~,
 \label{eq:Mijxbox}
\end{eqnarray}
\end{subequations}
where $p_1$, $p_2$, $p_3$, and $p_4$ are the momenta of the incoming quark, incoming anti-quark, outgoing positron, and outgoing electron, respectively, and
$D_{ij}$ is the  product of the denominators of the propagators in the loop; finally, $\bar{M}_1 = -M_1$ and $\bar{M}_2 = M_2$ (see Eq.~(\ref{eq:EigMass})).
The chirality projection operators in the Feynman amplitude
pick out the $\slashed{p}$ terms in the propagators of the standard box, and the mass terms in
those of the crossed box, which is also indicated by the mass insertions in the right diagram of Fig.~\ref{fig:AllTheFeynmen}.

In the summation in Eq.~(\ref{eq:pDiracamp}), the combinations with the same dark fermion in the upper and lower propagator $\mathcal{M}^{11}_{\text{box}} + \mathcal{M}^{11}_{\text{xbox}}$ and $\mathcal{M}^{22}_{\text{box}} + \mathcal{M}^{22}_{\text{xbox}}$,
are suppressed due to destructive interference as discussed above.
This leaves us with $(\mathcal{M}^{12}_{\text{box}} + \mathcal{M}^{12}_{\text{xbox}}) + (\mathcal{M}^{21}_{\text{box}} + \mathcal{M}^{21}_{\text{xbox}})$.

From Eq.~(\ref{eq:Mijxbox}), we see that the crossed box diagrams with two different dark fermions in the upper and lower propagator come with a relative minus sign with respect to the crossed box diagrams that contain only one dark fermion species (the numerators are $\bar{M}_1 \bar{M}_2 = - M_\chi(M_\chi + \Delta M)$, and $\bar M_1^2 = M_\chi^2$ or $\bar M_2^2 = (M_\chi + \Delta M)^2$, respectively). Therefore, $\mathcal{M}^{12}_{\text{box}} = \mathcal{M}^{21}_{\text{box}}$ interferes \emph{constructively} with $\mathcal{M}^{12}_{\text{xbox}} = \mathcal{M}^{21}_{\text{xbox}}$. Consequently, in the mixed dark matter case we expect that the monocline feature in the cross section
appears at a dilepton invariant mass of $ \MLL \simeq M_1 + M_2 = 2 M_\chi + \Delta M$.

\begin{figure}
\begin{center}
\includegraphics[width=8cm]{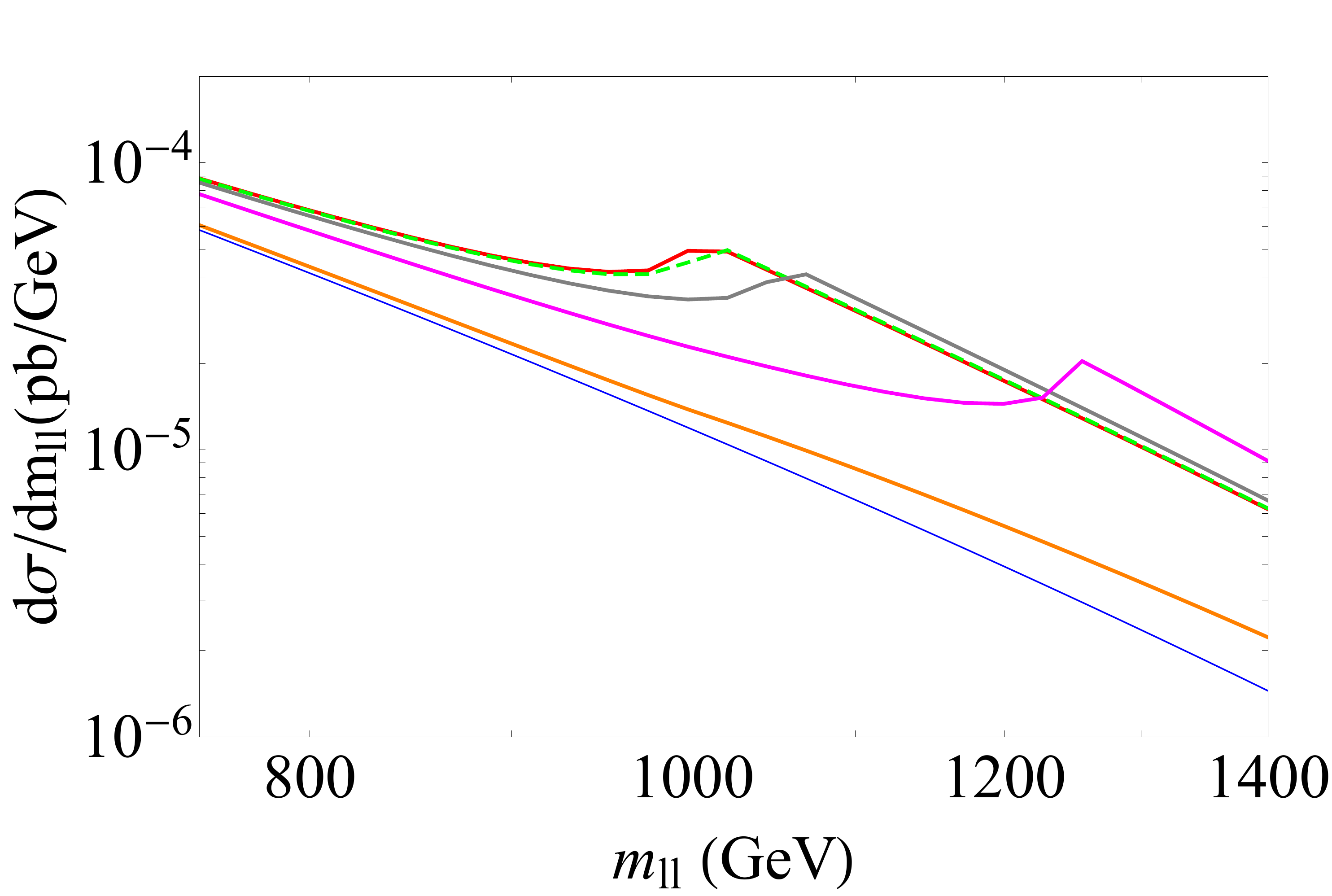}
\caption{The differential $pp \to \ell^+\ell^-$ cross section
as a function of the dilepton invariant mass
for a mixed dark matter particle
$\chi$, in \modu. Here, $\lambda=1.8$ and $\Mchi=M_\phi=500$ GeV. The color code
is -- blue: $\sigma_{\text{SM}}$, red: $\Delta M=0$ (pure Dirac), green dashed: $\Delta M=5$ GeV,
grey: $\Delta M=50$ GeV, magenta: $\Delta M=200$ GeV, orange: $\Delta M \ra \infty$ (pure Majorana).
}
\label{fig:pDiracXS}
\end{center}
\end{figure}

The pure Dirac and Majorana limits discussed above can now be more readily understood. When $\Delta M = 0$ (Dirac limit), the monocline feature appears at $\MLL \simeq 2 M_\chi $, as seen in Fig.~\ref{fig:DiracXS}. When $\Delta M \ra \infty$ (Majorana limit), the monocline feature is at $\MLL \ra \infty $ and is not observed.
As an illustration, we provide in Fig.~\ref{fig:pDiracXS} the dilepton invariant mass distribution in
\modu~with $\lambda = 1.8$ and $M_\chi = M_\phi = 500$~GeV for
intermediate values of the dark fermion mass splitting $\Delta M$ = 5, 50 and 200~GeV, given by green dashed, grey and magenta curves respectively.
The monocline is featured at $m_{\ell\ell} \simeq 1005, 1050$ and 1200~GeV, respectively. For comparison, Fig.~\ref{fig:pDiracXS} also shows the pure Dirac (red) and pure Majorana limits (orange).
We observe that a splitting of $\Delta M = 5$~GeV, a value that corresponds to
the pseudo-Dirac case, results in nearly identical behavior to that of pure Dirac dark matter.
To summarize, introducing two Weyl fields in the dark matter sector
     with the two eigenstates split by
    a small mass -- a scenario called pseudo-Dirac dark matter -- can give
    a dilepton invariant mass distribution that has almost exactly the same features as a pure Dirac
    dark matter particle in dilepton production.

\begin{figure}[t]
\begin{center}
\includegraphics[width=7cm]{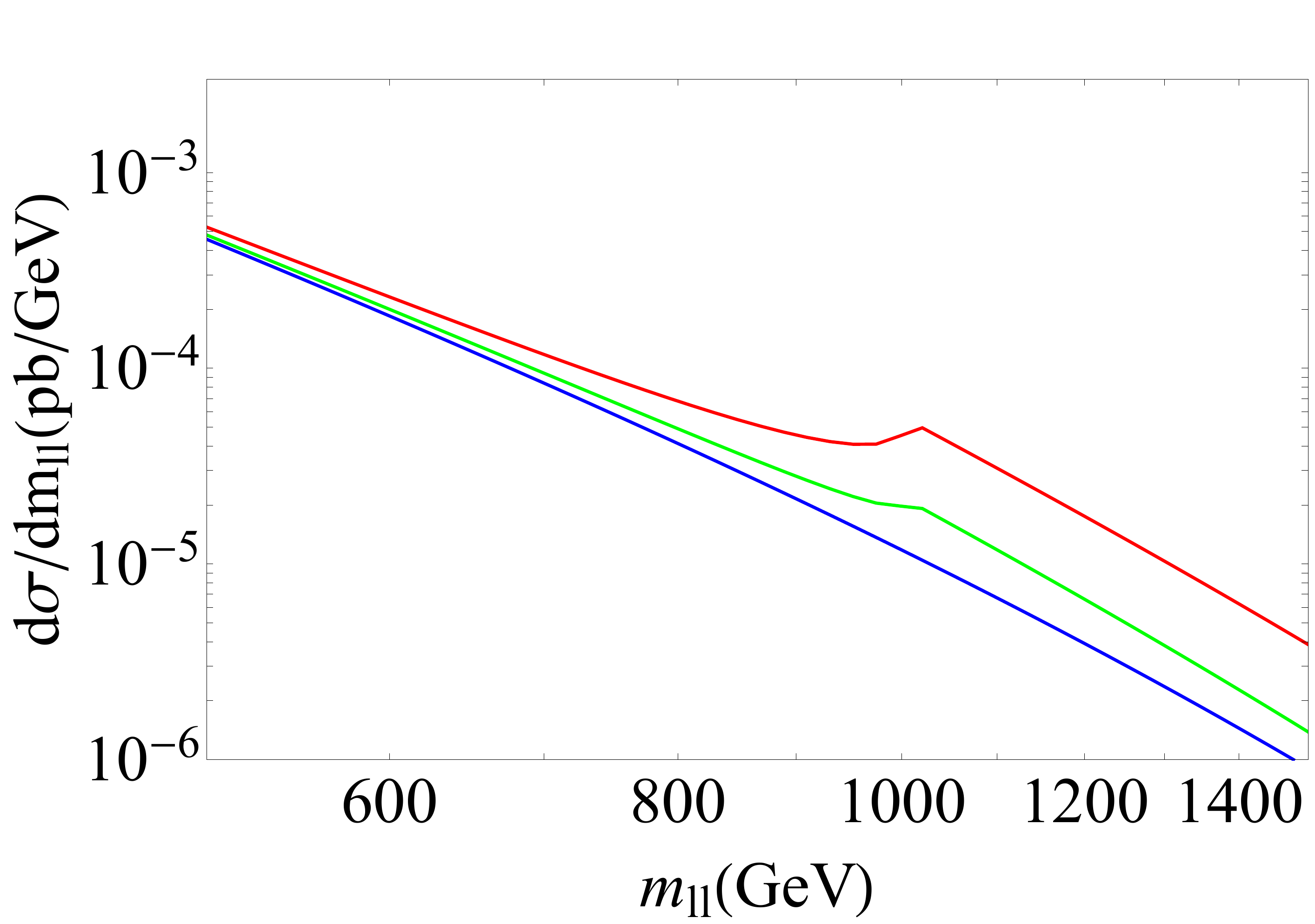}
\end{center}
\caption{The differential $pp \to \ell^+\ell^-$ cross section as a function of the dilepton invariant mass in \modu. Shown here are the effects of variation in $\lambda$ keeping $\Mchi = 500$~GeV fixed, with $M_\phi = \Mchi$ and $\Delta M = 0$. Here, red: $\lambda=1.8$, green: $\lambda=1.4$, blue: SM. }
\label{fig:pDiracXSCoupsVaried}
\end{figure}
\begin{figure}[t]
\begin{center}
\includegraphics[width=6.8cm,height=6.7cm]{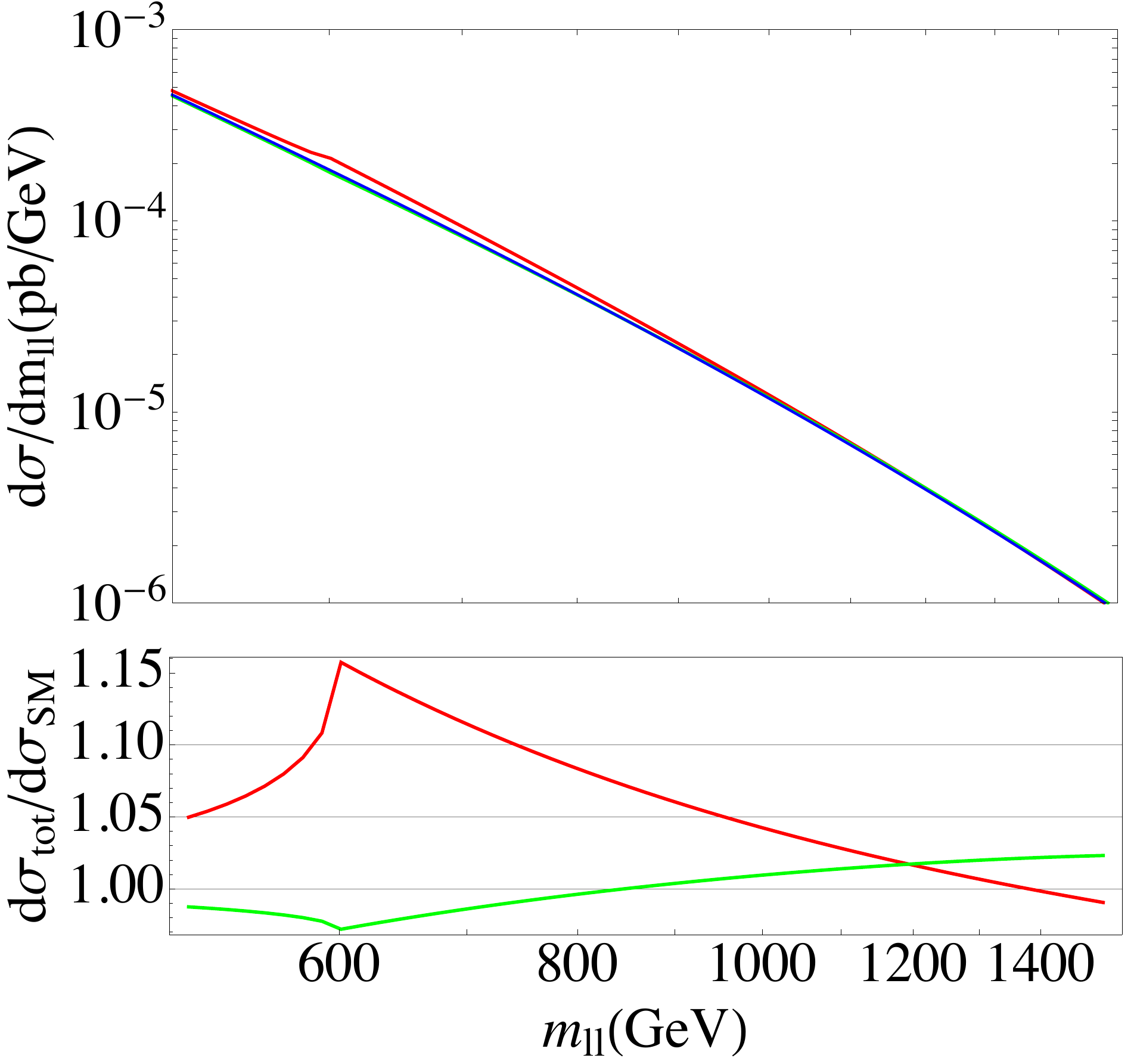}
\end{center}
\caption{Upper plot: The differential $pp \to \ell^+\ell^-$ cross section as a function of the dilepton invariant mass.  Lower plot: the ratio $d\sigma_{\text{tot}}/d\sigma_{\text{SM}}$ as a function of $\MLL$. 
We set $\lambda = 1$ and $\Mchi = 300$~GeV, with $M_\phi = \Mchi$ and $\Delta M = 0$.
Here red: \modu, green: \modd, blue: SM.
}
\label{fig:InterferenceRegimeCombinedPlot}
\end{figure}
\begin{figure}[t]
\begin{center}
\includegraphics[width=7cm]{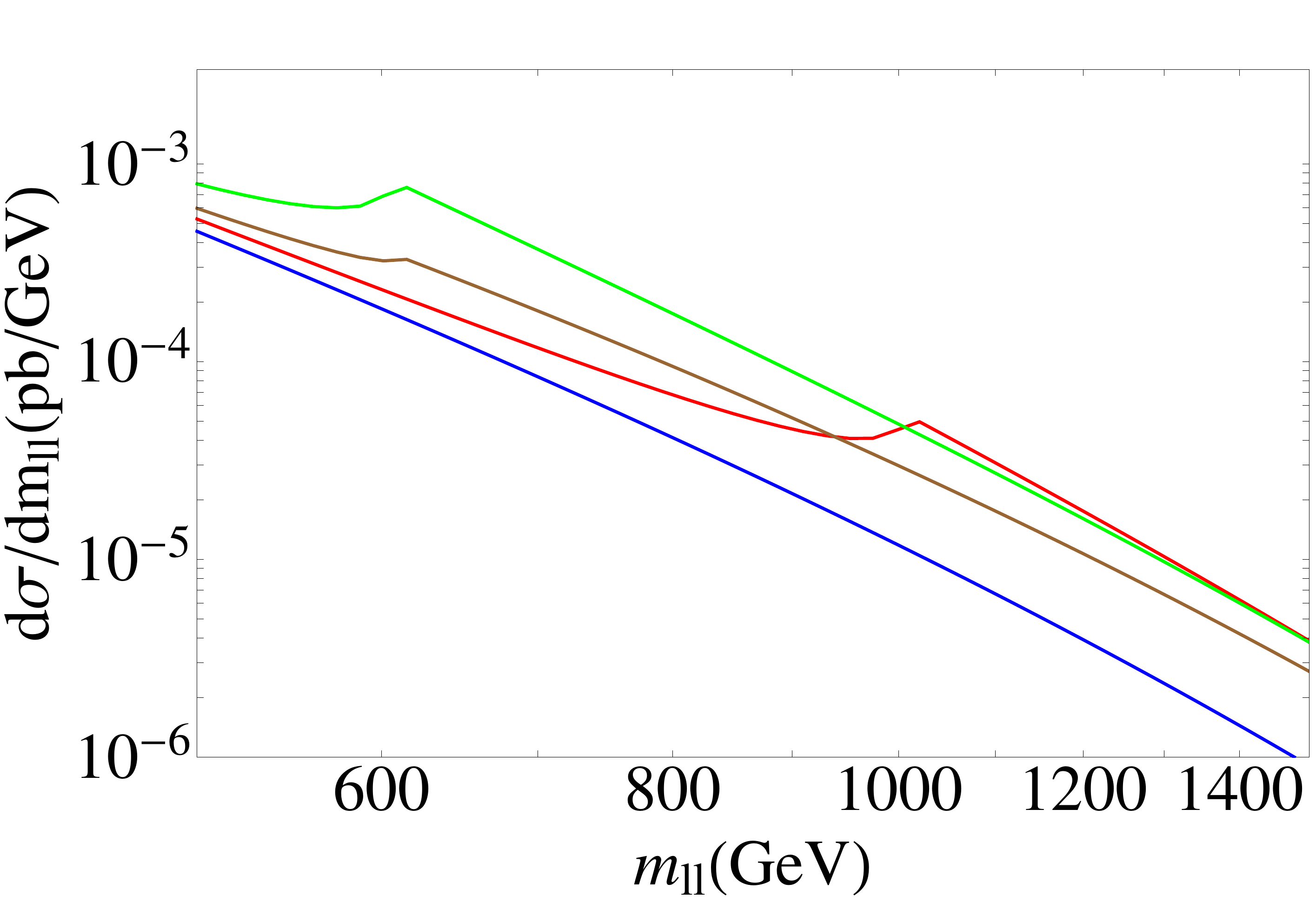}
\end{center}
\caption{The differential $pp \to \ell^+\ell^-$ cross section as a function of the dilepton invariant mass in \modu. Shown here are the effects of variation in $\Mchi$ and $M_\phi$ holding $\lambda = 1.8$ fixed and $\Delta M = 0$. 
Here, red: $M_\chi=M_\phi=500$~GeV, green: $M_\chi=M_\phi=300$~GeV, 
brown: $M_\chi=300$~GeV, $M_\phi=500$~GeV, blue: SM.}
\label{fig:pDiracXSMassesVaried}
\end{figure}

We end this section by discussing aspects of the dependence of the monocline feature on the mediator coupling $\lambda$ and the dark matter mass.
  The change in the size of the monocline feature for several values of the coupling
  $\lambda$ is shown in Fig.~\ref{fig:pDiracXSCoupsVaried}, where we fix
  $\Mchi$ = $500$ GeV and $\Delta M = 0$. The red curve corresponds
  to $\lambda=1.8$ and the green curve to $\lambda=1.4$, with the
   blue curve depicting the Standard Model LO value. As one would expect,
  the deviations from SM become less significant as the coupling is decreased.
  
  In Fig.~\ref{fig:InterferenceRegimeCombinedPlot}, we show the behavior of dilepton spectrum
  in the regime where the new physics signal is dominated by the interference term $d\sigma_{\text{int}}$. For illustration, we
  have taken $\lambda = 1$ and $\Mchi = 300$~GeV.  The upper and lower plots indicate
  the distribution $d\sigma/d\MLL$ and the ratio $d\sigma_{\text{tot}}/d\sigma_{\text{SM}}$
  respectively. The red and green curves in both plots represent \modu~and D
  respectively, with the blue curve in the upper plot denoting the SM at LO. 
  As expected, due to the smaller couplings the new physics effect on the dilepton rate is much smaller.
  Also seen are the interesting effects of destructive interference with
  the SM amplitude. In \modu, we see a reduction of the dilepton rate with
  respect to the SM ($d\sigma_{\text{tot}}/d\sigma_{\text{SM}} < 1$) for invariant masses considerably above the kinematic threshold $\MLL \gsim 1300$~GeV. In Model D on the other hand, destructive interference is present below and near the threshold, while for large invariant masses $\MLL \gsim 850$~GeV, the interference becomes again constructive. Note that in both models the new physics amplitudes have the same sign, while the sign of the SM amplitude differs due to the differing electric charge of the initial state quarks.
  
The variation of the signature as a function of the masses is seen in
Fig.~\ref{fig:pDiracXSMassesVaried},
where the coupling is fixed at $\lambda=1.8$ and $\Delta M = 0$.
The green and red curves take $M_\chi=M_\phi$ for two values, $300$ and 
$500$~GeV respectively.  
Notice that even though the location of the monocline is different for
different $\Mchi$'s, the size of the
deviation from the Standard Model is approximately independent 
of $M_\chi$. This is because when the new physics contribution 
is dominated by $|\mathcal{M}_{\text{box}}|^2$ and
$|\mathcal{M}_{\text{xbox}}|^2$,  as is the case for $\lambda \gsim 1.4$, 
for a fixed ratio of mediator to dark matter mass, $\Mrat$, the ratio
$d\sigma_{\text{total}}/d\sigma_{\text{LO}}$
is determined mainly by the coupling.
We also show the effect of splitting $M_\phi$ from $M_\chi$ in the
brown curve.  Notice that the sharp monocline rise is less pronounced 
near $\MLL = 2 M_\chi$ (compared with the green curve), and the 
size of the effect for $\sqrt{\hat{s}} > 2 M_\phi$ 
slowly asymptotes to the green and red curves.

\subsection{Angular Distribution}
\label{subsec:angdist}

The loop corrections in the model also leave their imprint 
in the angular distribution of the rates $d^2\sigma/d\MLL d\ct$, where
 $\ct \equiv \cos\theta$ with the angle $\theta$ already introduced in Eq.~(\ref{eq:sigma_tot}).
 In general, the angular distribution $d^2\sigma/d\MLL d\ct$ can be written as
\begin{equation}
 \frac{d^2\sigma}{d\MLL d\ct} = \sum_{n=0}^{\infty} a_n \ct^n~,~~~~ a_n\in\mathcal{R} ~.
 \label{eq:gendXSdct}
\end{equation}
In general, the $a_n$ coefficients are functions of $m_{ll}$.
 For an $s$-channel--mediated process (including the SM Drell-Yan process at
 tree level),  $a_0 = a_2$
 and $a_{n\geq3} = 0$. 
 Hence we can write \cite{Chatrchyan:2012dc}
\begin{equation}
\frac{d^2\sigma_{s-\text{chan.}}}{d\MLL d\ct}   \propto \frac{3}{8}(1+\ct^2) + A_\text{FB}(\MLL) \ \ct ~,
\label{eq:schandoublediff} \end{equation}
where $A_\text{FB}(\MLL)$ is the forward-backward asymmetry.
Therefore, the measurement of the forward-backward asymmetry 
$A_\text{FB}(\MLL)$ characterizes the
shape of the differential distribution for an
$s$-channel--mediated process. For a general distribution as given
in Eq.~(\ref{eq:gendXSdct}), more observables must be measured to
determine the coefficients~${a_n}$. 

The forward-backward asymmetry can be formally obtained as:
\begin{eqnarray}
 A_\text{FB}(\MLL) &\equiv& \frac{\int_{0}^{1} d\ct(d^2\sigma/d\MLL d\ct)  - \int_{-1}^{0} d\ct(d^2\sigma/d\MLL d\ct)}{\int_{-1}^{1} d\ct
 (d^2\sigma/d\MLL d\ct)}  \nonumber \\
 &=& \frac{(d\sigma/d\MLL)_F - (d\sigma/d\MLL)_B}{(d\sigma/d\MLL)_{\text{tot}}} \, . 
\label{eq:afbdef}
\end{eqnarray}

\begin{figure*}
\begin{center}
\includegraphics[width=7cm]{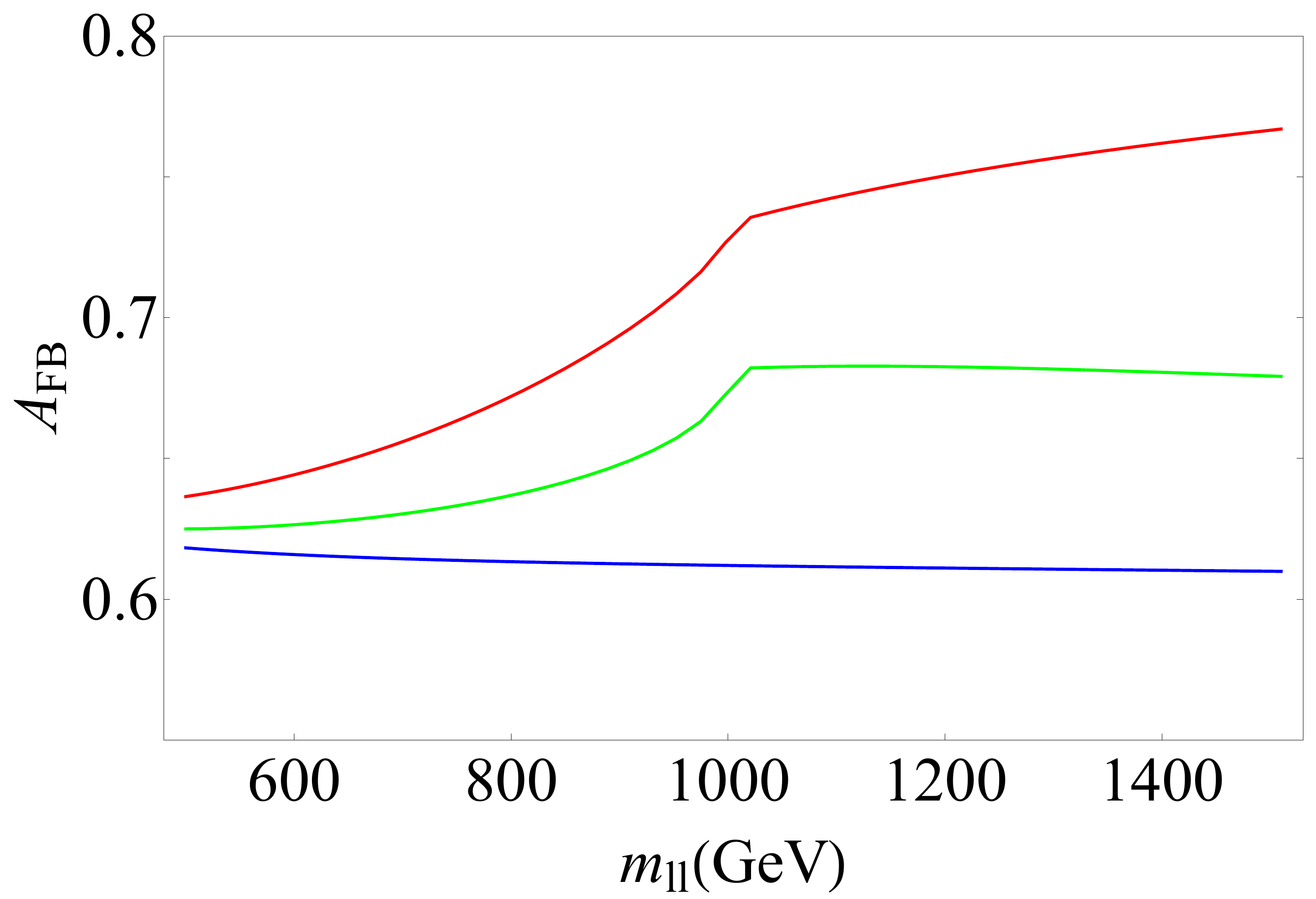}
\quad \quad \quad
\includegraphics[width=7cm]{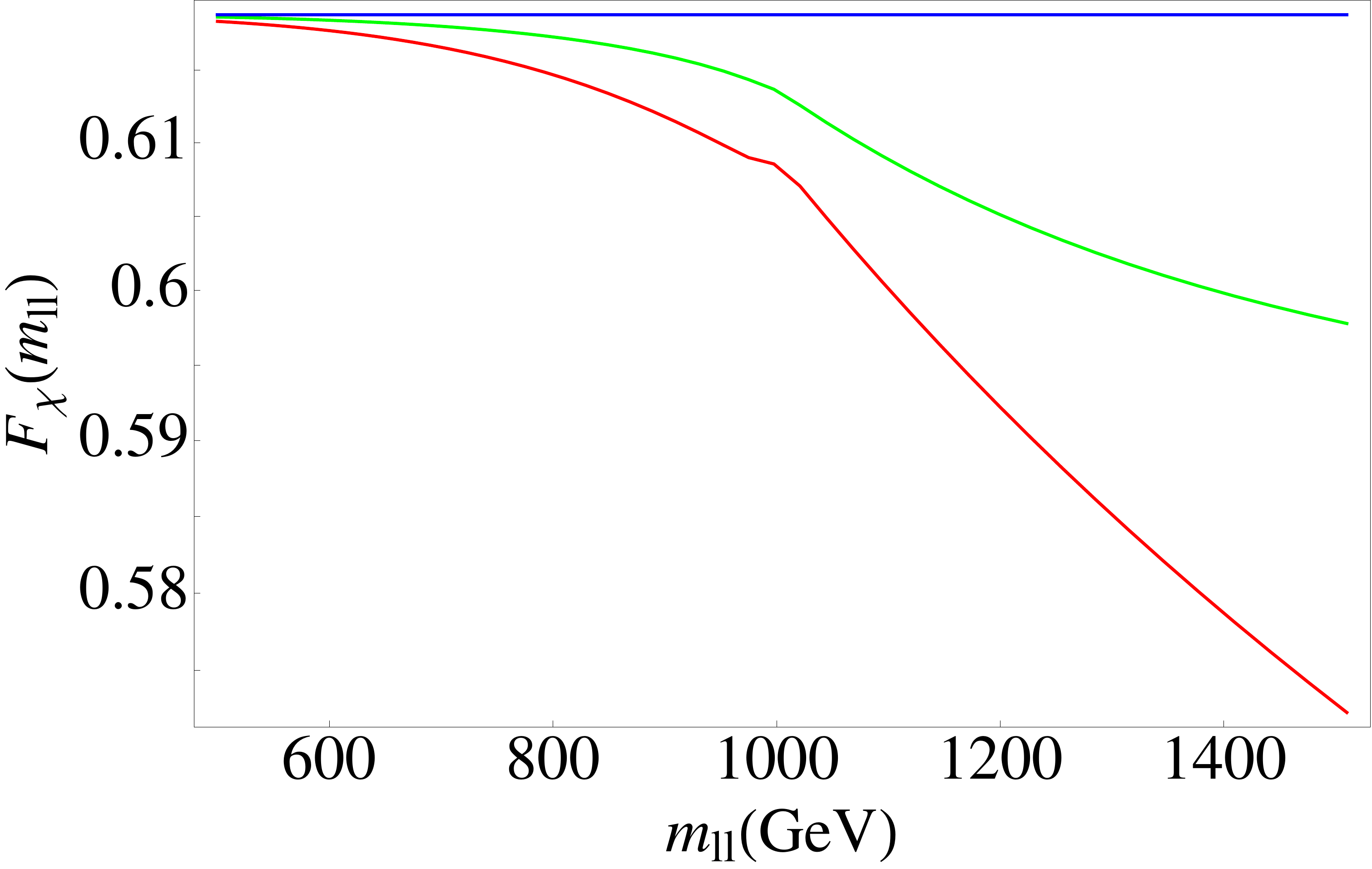}
\end{center}
\caption{LEFT: The forward-backward asymmetry as defined
in Eq.~(\ref{eq:afbdef}) as a function of the dilepton invariant mass; RIGHT: $F_\chi (m_{\ell \ell})$ as defined in Eq.~(\ref{eq:rcosthetadef}) as a function of the dilepton invariant mass.
In both plots, \modu~is used and masses are set to $M_\chi=M_\phi=500$ GeV and $\Delta M = 0$. Here, blue: Standard Model at LO ($\lambda$ = 0), green: $\lambda$ = 1.4,
 red: $\lambda$ = 1.8. All curves are commuted at the the partonic level.}
\label{fig:Angulars}
\end{figure*}

    The $A_{\text{FB}}(\MLL)$ computed at \emph{partonic} level
   in \modu~is illustrated in the plot on the left-hand side of Fig.~\ref{fig:Angulars}.
    The red curve corresponds to $\lambda = 1.8$ and the green curve to $\lambda=1.4$, with
    $\Mchi=M_\phi=500$ GeV with $\Delta M=0$ for both curves. The blue line denotes the Standard Model prediction at LO\@.
   We notice a significant increase of $A_\text{FB}$ at
   the threshold, which is the result of three different effects
   at $\MLL \simeq 2M_{\chi}$:
\begin{itemize}
 \item[(a)] an increase in
$[(d\sigma/d\MLL)_F^{\text{box}}$ $-$ $(d\sigma/d\MLL)_B^{\text{box}}]/$
$(d\sigma/d\MLL)_\text{tot}$
   due to a huge increase in $\sigma^{\text{Fwd}}_{\text{box}}$,
 \item[(b)] a slight increase in
$[(d\sigma/d\MLL)_F^{\text{int}}$
$-$ $(d\sigma/d\MLL)_B^{\text{int}}]/$ $(d\sigma/d\MLL)_\text{tot}$,
   and
 \item[(c)] a decrease in
   $[(d\sigma/d\MLL)_F^{\text{LO}}$ 
$-$ $(d\sigma/d\MLL)_B^{\text{LO}}]/$ 
$(d\sigma/d\MLL)_\text{tot}$, 
due to the increase in $(d\sigma/d\MLL)_\text{tot}$.
\end{itemize}
A search for new physics in dilepton production using $A_\text{FB}$ has been carried out by the ATLAS collaboration in~\cite{Aad:2014wca} using $20$ fb$^{-1}$ of 8~TeV data.
Due to the inherent uncertainties in the direction of the initial (anti)quark and the transverse momenta of the partons in a proton-proton collider, events are reconstructed by first boosting along a longitudinal direction and identifying the dilepton center-of-momentum frame. The quark, due to its predominantly valence nature, is then assumed to have originated in the direction of the boost. The details of constructing the angle of scattering $\theta^*$ in this so-called Collins-Soper (CS) frame~\cite{Collins:1977iv} are provided in~\cite{Aad:2014wca}. The inevitable misidentification of quarks (antiquarks) that comes with this procedure leads to ``mistagging'' a fraction of forward (backward) events as backward (forward), thus diluting the asymmetry. Higher order QCD corrections to the differential Standard Model cross section further symmetrize the forward-backward events.
As a result, the $\MLL$-dependent Standard Model values for $A_\text{FB}$ shown in~\cite{Aad:2014wca} are smaller than the ones in Fig.~\ref{fig:Angulars} by a factor of 1.5 -- 3.
A full fledged angular analysis that uses the CS frame and takes into account higher order corrections is beyond the scope of this work. 

A complementary way to probe the angular distribution are observables that quantify the preference of dilepton events in a predefined central region of the detector over events in the outer region. Measuring such observables does not require knowledge of the direction of the initial parton, making them potentially advantageous at a proton-proton collider. An example of this is the ATLAS measurement of the observable $F_\chi (m_{jj})$ in dijet distributions
at $\sqrt{s} = 7$ TeV~\cite{ATLAS:2012pu}. It is defined as $F_\chi \equiv N_\text{central}/ N_\text{total}$, where $N_\text{total}$ is the total number of events, and $N_\text{central}$ is the number of dijet events in a central region defined by $\chi \equiv \exp (2|y|) < \chi_\text{max}$, where $y$ is the rapidity of each jet in the dijet CM frame. In the ATLAS analysis, the observable $F_\chi$ is used to distinguish between isotropic new physics processes and QCD backgrounds, that prefer the forward direction. As a simple illustration of their applicability to our model, we compute the quantity,
\begin{equation}
F_\chi (m_{\ell \ell}) \equiv  \frac{\int_{-a}^a (d\sigma/dc_\theta) dc_\theta}{\int_{-1}^{1} (d\sigma/dc_\theta) dc_\theta}~,
\label{eq:rcosthetadef}
\end{equation}
where the central region is defined by $-a \leq c_\theta \leq a$. Choosing $a=1/2$ (which corresponds to $\chi = 3$), we plot $F_\chi (m_{\ell \ell})$ at the partonic level in \modu~with
$\Mchi = M_\phi = 500$~GeV and $\Delta M=0$ on the right-hand side of Fig.~\ref{fig:Angulars}. The red curve corresponds to $\lambda = 1.8$, the green curve to $\lambda=1.4$, and the blue curve depicts the Standard Model at LO.

The SM curve appears flat 
which can be understood as follows. One sees from Eq.~(\ref{eq:schandoublediff}) that for a given $\MLL$, the angular distribution can be written as
 $d\sigma/dc_\theta (\MLL, c_\theta) = f(\MLL)[\frac{3}{8}(1+\ct^2) + A_\text{FB}(\MLL) \ct]$. From the left-hand plot in Fig.~\ref{fig:Angulars}, we see that in the SM $A_\text{FB}$ is largely insensitive to $\MLL$ for the range considered because all SM states can be taken as massless for this range and there is no mass scale in the problem. 
Thus $d\sigma/dc_\theta (\MLL, c_\theta)$ can be
 approximately written as $f(\MLL)[\frac{3}{8}(1+\ct^2) + A_\text{FB} \ct]$. Therefore, to a good approximation, $f(\MLL)$ drops out of $F_\chi (m_{\ell \ell})$. In general no such approximate factorization can be made for the new physics effects in our model. 
We find that the new physics box amplitude tends to slightly favor the outer regions over the central region. The values for $F_\chi (m_{\ell \ell})$ in our model are therefore always smaller than the Standard Model's unless interference effects lead to a deficit in rates with respect to the SM. The preference for the outer regions gets more pronounced for $\MLL \gsim 2 \Mchi$, where also the imaginary part in the amplitude turns on. This behavior is reflected in the red curve by a kink at $\sim 1000$ GeV on the right-hand-side plot in Fig.~\ref{fig:Angulars}, to the right of which the distribution falls steeper.

\subsection{Dilepton Spectrum Constraints}

We can compare
the predicted dilepton spectra of our model to measurements by the LHC collaborations \cite{Aad:2014cka,Chatrchyan:2012oaa}
by conducting a shape analysis.
The dominant Standard Model background in these searches is the Drell-Yan process,
which at tree-level is $s-$channel photon-- and $Z$--mediated as
 shown in Fig.~\ref{fig:AllTheFeynmen}.  Subdominant backgrounds come from the production
 of tops, dibosons, dijets and W+jet.
Both
ATLAS and CMS find their observed dilepton spectra are consistent 
with the Standard Model.

ATLAS has dilepton events with invariant masses as high as
$\sim 1600$ GeV (1800 GeV) for $m_{ee}$ ($m_{\mu\mu}$), whereas CMS has events up to 
$\sim 1750$ GeV (1850 GeV).
The ATLAS and CMS measurements can be translated into constraints of our model.
In our
 analysis we only consider $\MLL$ bins that are far from the $Z$-resonance
given the dark fermion masses we consider.
In order to generate signal spectra,
we first analytically compute 
the $pp \to \ell^+\ell^-$ cross section ratios $d\sigma_{\text{total}}/d\sigma_{\text{SM}}$
bin by bin using the MSTW2008NNLO parton distribution functions, where
$d\sigma_{\text{total}}$ and $d\sigma_{\text{SM}}$ are as defined in Eq.~(\ref{eq:XStotalDirac}).
We choose the squared factorization scale and $Q^2$ to be $m_{ll}^2$.
We then scale
the experimentally provided Drell-Yan NNLO backgrounds by these ratios. We do not
consider the subdominant backgrounds.

Bounds 
on the model parameter space 
can be set by comparing the dilepton spectra of our model with the Standard Model predictions, by computing
$\Delta \chi^2 = \chi^2_\text{NP} - \chi^2_\text{SM} $, where
\begin{eqnarray}
\chi^2_\text{NP} = \sum_{i=1}^{N_{\text{bins}}}  \frac{(N^i_{\text{obs}}-N^i_{\text{NP}})^2}{N^i_{\text{NP}}+\sigma^2_{\text{SM}}} ~, \\
\chi^2_\text{SM} = \sum_{i=1}^{N_{\text{bins}}}  \frac{(N^i_{\text{obs}}-N^i_{\text{SM}})^2} {N^i_{\text{SM}}+\sigma^2_{\text{SM}}} ~,
\label{eq:DeltaChiSq}
\end{eqnarray}
with $N^i_{\text{NP}}$ the number of events expected by our model,
$N^i_{\text{SM}}$ the number of events predicted by the SM,
$N^i_{\text{obs}}$ the number of events observed and $\sigma_{\text{SM}}$
is the background systematic uncertainty.
By setting
$\Delta \chi^2 = 5.99$, we obtain a 95\% C.L. exclusion limit in the $\lambda-\Mchi$
plane
with respect to the Standard Model.
In the following, we compare the model with the ATLAS results~\cite{Aad:2014cka}.
ATLAS and CMS have comparable sensitivities and their results are in good agreement with each other. Therefore, using the CMS results~\cite{Chatrchyan:2012oaa} would lead to very similar exclusion limits.
We do not attempt a statistical combination of the ATLAS and CMS results.

As one would expect, in general the shape of the dilepton spectrum is sensitive to $\lambda$.
For instance, depending on the Model (U, D or UD) used for
setting constraints, it is possible to obtain also a slight \textit{deficit} in model events with respect
to the background, due to interference effects for dilepton invariant masses below the kinematic threshold (see also~\cite{Bai:2014fkl} for a recent study of destructive interference effects at colliders.). This typically occurs at $\lambda \lsim 1$.
We will find, however, that our $\Delta \chi^2$ analysis at $\sqrt{s} = 8$ TeV is only sensitive to $\lambda \gsim 1.4$,
 where the
signal is dominated by $d\sigma_{\text{box}} \propto |\mathcal{M}_{\text{box}}|^2$.
Thus, the nature of the model spectrum is as discussed in Subsection~\ref{subsec:pDiracRates}.
It then follows that the largest contributions to $\Delta \chi^2$ comes from the contribution near $\MLL \simeq M_1 + M_2$, where the monocline feature leads to the largest signal over background. 

We will discuss the results of the $\chi^2$ analysis in 
Sec.~\ref{sec:constraintsummary} along with
additional constraints on our parameter space from dedicated dark matter searches at the LHC, from direct detection experiments and from the dark matter relic abundance.

\begin{figure*}
\begin{center}
\includegraphics[width=11.5cm]{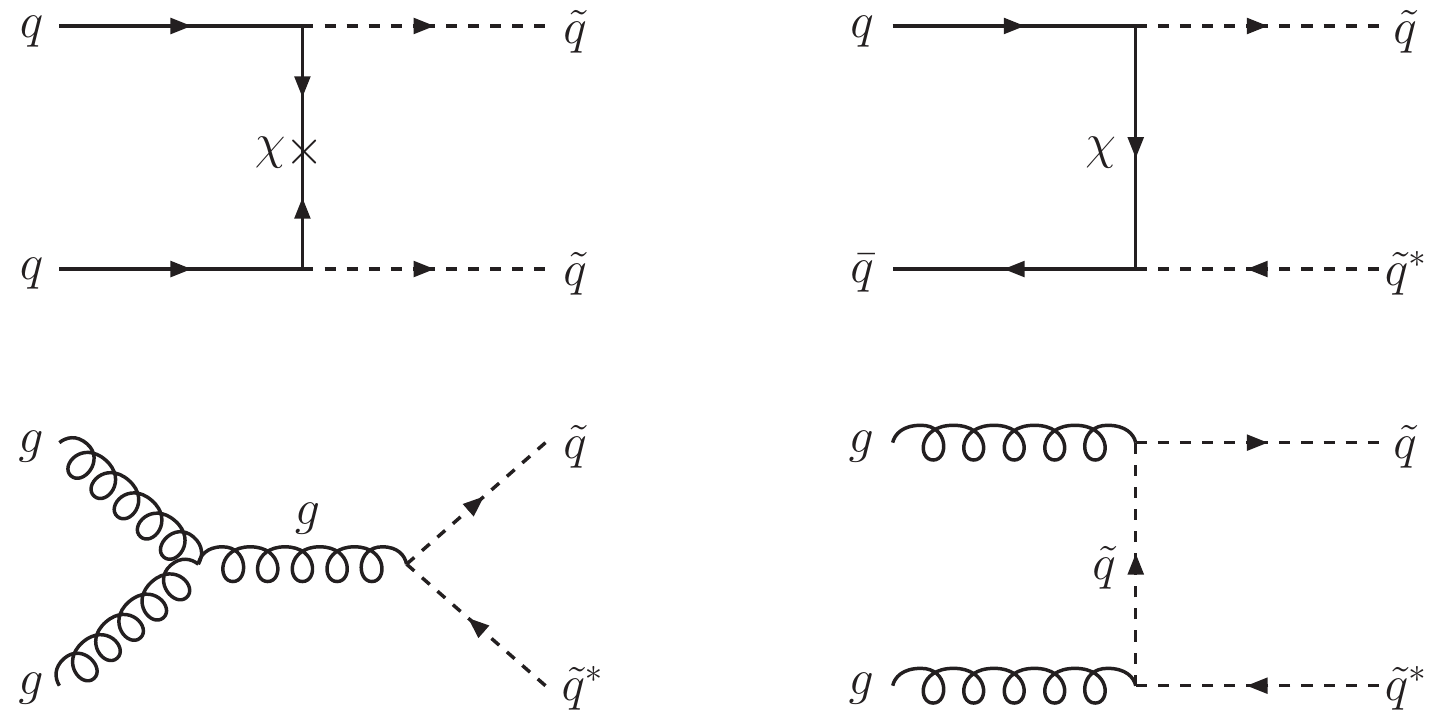}
\caption{Example Feynman diagrams with the highest contribution to the pair production of the colored mediator, resulting in jets+MET signals.}
\label{fig:FeynMJETs}
\end{center}
\end{figure*}

\section{Related Constraints}
\label{sec:constraints}

The primary focus of our paper is on the new signals of radiative
corrections of dark matter on the dilepton kinematical and
angular distributions.  There are, of course, several correlated
implications, from LHC predictions, the thermal relic density, to
the predictions for the scattering rates in direct detection 
experiments.  In this section we consider the \emph{constraints}
that these correlated implications place on the parameter space
of the simplified models that we consider.  
We consider the bounds set by jets + MET searches at the
LHC~\cite{Chatrchyan:2014lfa,Aad:2014wea},
the bounds from nucleon-dark matter scattering in 
direct detection experiments
\cite{Aprile:2012nq,Akerib:2013tjd,Aprile:2013doa}, 
and the dark matter thermal relic abundance 
(now best determined by Planck~\cite{Ade:2013zuv}).
Additional constraints can arise from the anomalous magnetic moment 
of the muon~\cite{Jegerlehner:2009ry} as well as from 
LEP results on four-lepton contact interactions~\cite{Schael:2013ita}.
In this section we step through each of these, 
detailing the various mechanisms behind each probe 
and how they place constraints on the model. A summary of all 
constraints and a comparison to the dilepton signal will be presented
in Sec.~\ref{sec:constraintsummary}.

\subsection{LHC constraints}

While searches for dark matter signals in the form of 
missing transverse energy (MET)+initial state radiation, 
the so-called mono-X signatures, are ongoing, the strongest constraints 
on our model come from recasted supersymmetry searches for
jets+MET signatures from ATLAS and CMS~\cite{Chatrchyan:2014lfa,Aad:2014wea}.
Indeed, pair production of the colored mediators, followed by the 
decay of the mediators into dark matter and a light quark contribute 
to the jets+MET signal. Some important
diagrams are shown in Fig.~\ref{fig:FeynMJETs}.
For recasting, we use the CMS T2qq simplified model
in~\cite{Chatrchyan:2014lfa}, where
the gluino is assumed decoupled and squark pair production is
followed by prompt decay
to a pair of LSPs with a branching ratio of 100\%.
Contours of the exclusion cross-sections in the plane of LSP mass and squark mass
are provided, which we compare with our signal cross-sections generated
at leading order using MadGraph5~\cite{Alwall:2011uj} with
CTEQ6L1 parton distribution functions~\cite{Pumplin:2002vw}. 
We will present the results of the numerical analysis in Sec.~\ref{sec:constraintsummary}, where we also compare the bounds with those obtained from the dilepton spectra.

Note that the supersymmetry search
assumes the squarks are pair-produced predominantly via
an $s$-channel gluon whereas the dominant production channel in our model is
$t$-channel exchange of $\chi_1$ and $\chi_2$. While in principle this leads to
different detector acceptances for the two processes, in practice we find
that these two acceptances are similar within a few percent, validating our use of the CMS
bounds for constraining our model.

   \subsection{Relic Abundance}
   \label{subsec:relic}

\begin{figure*}
\begin{center}
\includegraphics[width=11.5cm]{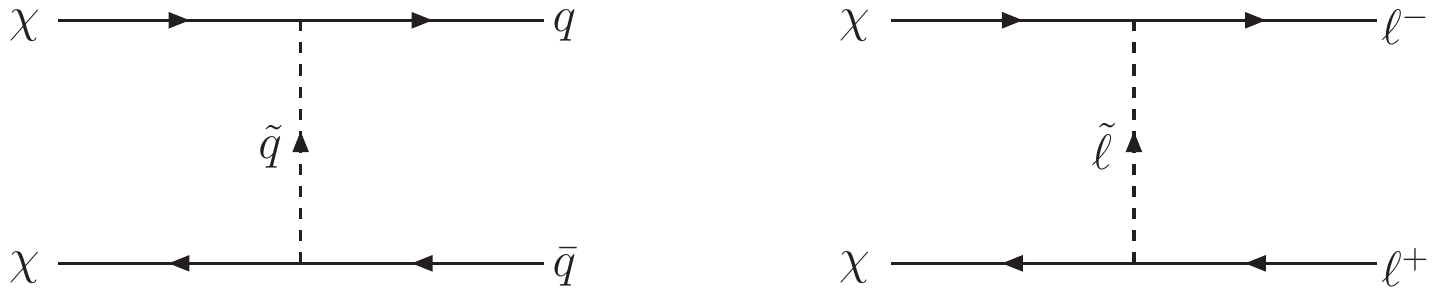}
\caption{Example dark matter annihilation diagrams that set the thermal relic abundance.}
\label{fig:FeynRelic}
\end{center}
\end{figure*}
\begin{figure*}
\begin{center}
\includegraphics[width=11.5cm]{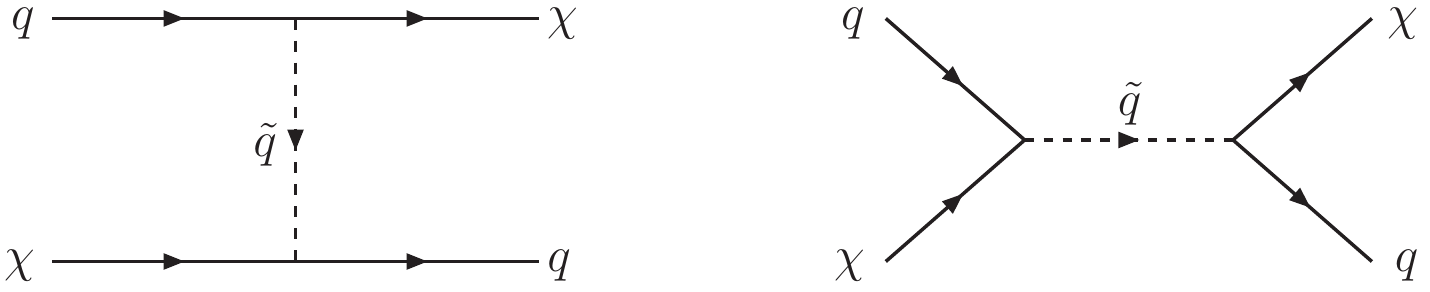}
\caption{Feynman diagrams contributing to direct detection signatures.}
\label{fig:FeynDD}
\end{center}
\end{figure*}

   If $\chi$ is a thermal relic of freeze-out,
   the diagrams in Fig.~\ref{fig:FeynRelic} contribute to its annihilation
   into SM fermions in the early universe.
    We can then calculate the relic abundance as a
   function of the masses and couplings in our model by solving
   the Boltzmann equation under the freeze-out condition. For the
   case of a pseudo-Dirac dark matter candidate (i.e. for small mass splitting between the two dark fermion states $\chi_1$ and $\chi_2$), coannihilations
   between the two eigenstates ($\chi_1 \chi_2 \ra
   f \bar{f}$ and $\chi_2 \chi_2 \ra
   f \bar{f}$)
    play an important role in setting the abundance.
    We incorporate
    these effects through an 
    effective cross-section \cite{Griest:1990kh} 
        \begin{equation}
        \sigma_{\text{eff}}(x) = \frac{\sigma_{11} + 2\sigma_{12}(1+\delta)^{3/2}e^{-x\delta}
    + \sigma_{22}(1+\delta)^{3}e^{-2x\delta}  }{(1+(1+\delta)^{3/2}e^{-x\delta})^2 } ~,
    \label{eq:effannihilationXS}
                \end{equation}
   where $x \equiv T/M_\chi$ is the ratio of temperature and dark matter mass and $\delta \equiv \delM$ is the fractional mass
   splitting between the dark matter states. For a splitting less than or comparable
   to the freeze-out temperature ($\delM  \lsim 1/x_F$), efficient $s-$wave annihilation
   of the $\sigma_{12}$ term in Eq.~(\ref{eq:effannihilationXS}) leads to small
   relic abundances that do not overproduce dark matter for large ranges of
   parameters in our model. For ($\delM \gg 1/x_F$), exponential suppression
   of the coannihilation terms in Eq.~(\ref{eq:effannihilationXS}) implies
   $\sigma_\text{{eff}} \approx \sigma_{11}$, whose $s-$wave component is
   chirality-suppressed by a factor of $(m_f/M_\chi)^2$. The dominant component in that case is p-wave suppressed, leading to larger relic abundances.
   While there is potentially a sub-dominant contribution from coannihilation between the
   scalar mediators and dark matter for $\Mrat \lsim 1.1$.  We neglect these effects
   in setting our bounds, since, as discussed below, we will find that constraints from direct
   detection are typically stronger than constraints from the dilepton spectrum
   in these regions of parameter space.

   The relic abundance is given by
   \begin{equation}
   \Omega_\chi h^2 \approx \frac{1.07 \times 10^9~\text{GeV}^{-1} }{M_\text{Pl}}\frac{x_F}{\sqrt{g_*}} \frac{1}{I_a+3I_b/x_F}  ~,
   \end{equation}
where the freezeout temperature $x_F$ can be determined through
   \begin{equation}
   e^{x_F} = \frac{5}{4}\sqrt{\frac{45}{8}}\frac{M_1 M_{\text{Pl}}(I_a + 6I_b/x_F)}{\pi^3\sqrt{g_*}\sqrt{x_F}} ~.
   \end{equation}
        The terms $I_a$ and $I_b$ quantify the integration over thermal history
        of the annihilating species before freeze-out, and are given by
   \begin{equation}
   I_a = x_F \int^\infty_{x_F} \frac{dx}{x^2}~a_{\text{eff}}~, ~~   I_b = 2 x_F^2 \int^\infty_{x_F} \frac{dx}{x^3}~b_{\text{eff}}~,
   \end{equation}
   where $\langle \sigma_{\text{eff}}v_{\text{rel}} \rangle = a_{\text{eff}} + b_{\text{eff}}v_{\text{rel}}^2$.
   Expressions for $a_{\text{eff}}$ and $b_{\text{eff}}$ in our model are given
   in Appendix~\ref{sec:aeffbeff}.
   
   The constraints on the model parameter space from the relic abundance will be shown in Sec.~\ref{sec:constraintsummary}.

   \subsection{Direct Detection}
   \label{subsec:dd}

   Dark matter is also constrained by underground experiments
   studying the recoil spectra of local galactic dark matter scattering off
   nuclei of heavy elements. Fig.~\ref{fig:FeynDD} shows the dominant diagrams in
   our model contributing to the scattering cross-section.

    The current best bounds for spin-independent scattering
   are set by results of the 85.3 day-run of the Large
   Underground Xenon (LUX) experiment~\cite{Akerib:2013tjd} and those for
   spin-dependent scattering by the XENON100 experiment~\cite{Aprile:2013doa}. The energy transfer in these
    scattering experiments is
    $\mathcal{O}(10 \; \text{keV})$, hence for a sufficiently large splitting 
    in the eigenmasses of pseudo-Dirac dark matter, only the
    lighter eigenstate takes part in the scattering. Such a scenario
    emulates a Majorana dark matter candidate scattering off the nucleus.
    In the Majorana case, the leading contribution to spin-independent scattering 
    comes from a quark twist-2 operator, which is suppressed by $1/M_\phi^8$.
    Therefore one only obtains modest bounds from spin-independent scattering.
    Constraints from spin-dependent scattering are typically comparable.
    
    On the other hand, in the pure Dirac limit ($\Delta M = 0$) and in the pseudo-Dirac case with a sufficiently small splitting in the dark fermion masses, the spin-independent scattering cross-section is dominated by the vector-vector interaction operator (which is absent for a Majorana fermion).
   These cases are subject to
   very stringent limits by spin-independent direct detection. Constraints from spin-dependent scattering of Dirac dark matter (where one finds a cross-section that is four times smaller than in the Majorana case) are however not relevant.
   
Following \cite{Chang:2013oia}, the direct detection cross sections
spin-independent scattering $\sigma_{\text{SI}}$ and spin-dependent 
scattering $\sigma_{\text{SD}}$  predicted by our model can be 
calculated using the formulae given in Appendix~\ref{sec:ddformulae}.

   \subsection{LEP Constraints}
   \label{subsec:LEP}

LEP analyses of four lepton contact interactions that contribute to $e^+e^- \to \ell^+\ell^-$ can also be used to place constraints on the parameter space of our model. Box diagrams with dark fermions and lepton mediators will generate four fermion interactions of the type $(\bar e \gamma_\mu P_R e)^2$. However, in agreement with~\cite{Freitas:2014jla} we find that the LEP results collected in~\cite{Schael:2013ita} give only mild constraints on our scenario. In particular, couplings $\lambda \lesssim 2$ are only constrained for very light dark matter masses of $M_\chi \lesssim 250$~GeV.

   \subsection{Anomalous Magnetic Moment of the Muon}
   \label{subsec:g-2}


One additional constraint in the case the dark fermions of our model interact with muons comes in principle from the anomalous magnetic moment of the muon, $(g-2)_\mu$. Indeed, loops with dark fermions and scalar mediators can contribute to $(g-2)_\mu$. The sign of the contribution to $(g-2)_\mu$ is fixed, and turns out to \emph{increase} the longstanding discrepancy of the observed value with respect to the theory prediction~\cite{Jegerlehner:2009ry,Agrawal:2014ufa}. Requiring that the model prediction for $(g-2)_\mu$ does not deviate by more than $5\sigma$ from the measured value, we find constraints only in extreme corners of parameter space with $\lambda_\mu \gtrsim 2$ and $M_\chi \sim M_\phi \lesssim 200$~GeV.

\section{Summary of all Constraints}
\label{sec:constraintsummary}

\begin{figure*}
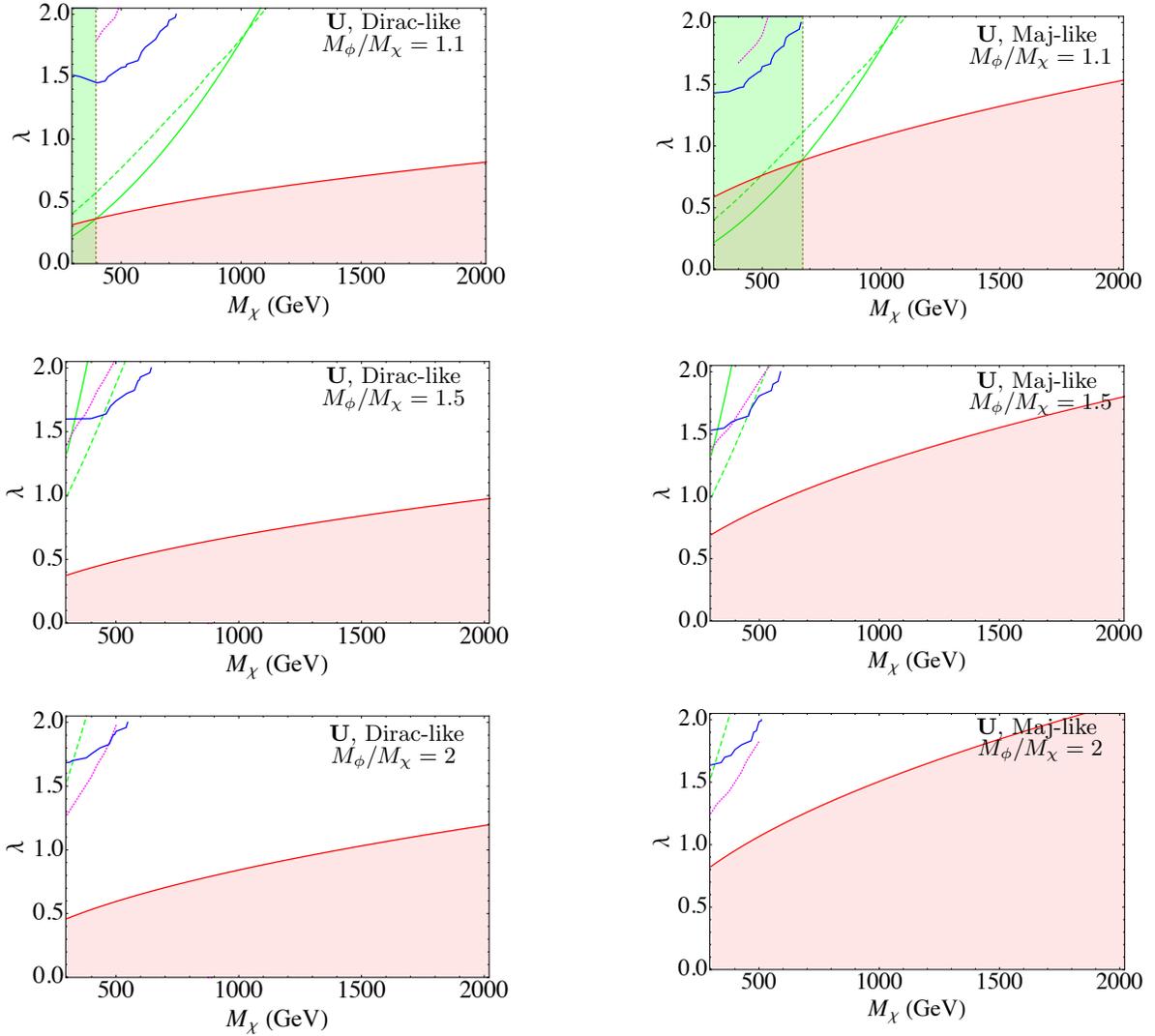

\begin{minipage}{0.49\textwidth}
\begin{lpic}[l(1mm)]{ConstraintsAlluPt1ade(.24,.24)}
\lbl[t]{225,588;\textbf{U}, Dirac-like}
\lbl[t]{225,576;$\Mrat = 1.1$}
\lbl[t]{225,385;\textbf{U}, Dirac-like}
\lbl[t]{225,373;$\Mrat = 1.5$}
\lbl[t]{227,180;\textbf{U}, Dirac-like}
\lbl[t]{225,168;$\Mrat = 2$}
\end{lpic}
\end{minipage}
\begin{minipage}{0.49\textwidth}
\begin{lpic}[r(1mm)]{ConstraintsAlluPt5ade(.23,.23)}
\lbl[t]{233,600;\textbf{U}, Maj-like}
\lbl[t]{235,588;$\Mrat = 1.1$}
\lbl[t]{233,393;\textbf{U}, Maj-like}
\lbl[t]{235,381;$\Mrat = 1.5$}
\lbl[t]{233,186;\textbf{U}, Maj-like}
\lbl[t]{235,174;$\Mrat = 2$}
\end{lpic}
\end{minipage}
\caption{Constraints in the plane of dark matter mass $M_\chi$ vs. coupling $\lambda$ in \modu. LEFT: $\delM = 0.1$ (Dirac-like at freeze-out), RIGHT: $\delM=0.5$ (Majorana-like at freezeout). The first, second and third rows correspond, respectively, to $\Mrat$ = 1.1, 1.5 and 2. The blue curve indicates the 95\% C.L. exclusion limit from the dilepton signal; the region above the dotted magenta curve is excluded by the CMS jets+MET search; the solid and dashed green curves represent the spin-independent LUX bounds and spin-dependent XENON100 bounds respectively; $\Omega_\chi h^2 > 0.12$ in the red shaded region. The green shaded regions, that are to the left of the intersections between the green and red curves, correspond to regions of parameter space that are excluded by direct detection experiments, even taking into account the reduced local relic density as predicted in these regions of parameter space.}
\label{fig:AllConstraintsu}
\end{figure*}

\begin{figure*}
\begin{minipage}{0.49\textwidth}
\begin{lpic}[l(-2mm)]{ConstraintsAlldPt1ade(.24,.24)}
\lbl[t]{250,617;\textbf{D}, Dirac-like}
\lbl[t]{250,605;$\Mrat = 1.1$}
\lbl[t]{250,405;\textbf{D}, Dirac-like}
\lbl[t]{250,393;$\Mrat = 1.5$}
\lbl[t]{250,192;\textbf{D}, Dirac-like}
\lbl[t]{250,180;$\Mrat = 2$}
\end{lpic}
\end{minipage}
\begin{minipage}{0.49\textwidth}
\begin{lpic}[r(-2mm)]{ConstraintsAlldPt5ade(.24,.24)}
\lbl[t]{247,613;\textbf{D}, Maj-like}
\lbl[t]{250,601;$\Mrat = 1.1$}
\lbl[t]{247,403;\textbf{D}, Maj-like}
\lbl[t]{250,391;$\Mrat = 1.5$}
\lbl[t]{250,192;\textbf{D}, Maj-like}
\lbl[t]{250,180;$\Mrat = 2$}
\end{lpic}
\end{minipage}
\caption{Constraints in the plane of dark matter mass $M_\chi$ vs. coupling $\lambda$ in \modd.
Plots and color coding as in Fig.~\ref{fig:AllConstraintsu}.}
\label{fig:AllConstraintsd}
\end{figure*}

\begin{figure*}
\begin{minipage}{0.49\textwidth}
\begin{lpic}[l(-2mm)]{ConstraintsAlludPt1ade(.24,.24)}
\lbl[t]{250,622;\textbf{UD}, Dirac-like}
\lbl[t]{250,610;$\Mrat = 1.1$}
\lbl[t]{250,405;\textbf{UD}, Dirac-like}
\lbl[t]{250,393;$\Mrat = 1.5$}
\lbl[t]{250,192;\textbf{UD}, Dirac-like}
\lbl[t]{250,180;$\Mrat = 2$}
\end{lpic}
\end{minipage}
\begin{minipage}{0.49\textwidth}
\begin{lpic}[r(-2mm)]{ConstraintsAlludPt5ade(.235,.235)}
\lbl[t]{254,630;\textbf{UD}, Maj-like}
\lbl[t]{254,618;$\Mrat = 1.1$}
\lbl[t]{254,415;\textbf{UD}, Maj-like}
\lbl[t]{254,403;$\Mrat = 1.5$}
\lbl[t]{254,198;\textbf{UD}, Maj-like}
\lbl[t]{254,186;$\Mrat = 2$}
\end{lpic}
\end{minipage}
\caption{Constraints in the plane of dark matter mass $M_\chi$ vs. coupling $\lambda$ in \modud.
Plots and color coding as in Fig.~\ref{fig:AllConstraintsu}.}
\label{fig:AllConstraintsud}
\end{figure*}

We can now combine all the constraints discussed in the sections above.
Figs.~\ref{fig:AllConstraintsu}, \ref{fig:AllConstraintsd} and \ref{fig:AllConstraintsud}
depict the regions of parameter space in the plane of mediator coupling $\lambda$ and dark matter mass $\Mchi$ that are allowed by all
experimental bounds for Models U, D and UD respectively. In all three figures, the
plots on the left-hand side correspond to a mass splitting between the dark matter states of
$\delM = 0.1$ (to represent a $\chi$ that is
Dirac-like at freeze-out). The plots on the right-hand side correspond to $\delM = 0.5$ (Majorana-like
at freeze-out). The rows correspond to ratios of mediator mass to dark matter mass of $\Mrat = 1.1, 1.5$ and 2 respectively.
The blue curves show the $95\%$~C.L. exclusion limit for a comparison between our model and the
 dilepton spectrum measured by the ATLAS collaboration at 8 TeV \cite{Aad:2014cka}. The dotted magenta
 curves depict the jets+MET bounds, recast from the CMS search for supersymmetry \cite{Chatrchyan:2014lfa}.
In the shaded red region the model overcloses the universe at freeze-out, with $\Omega_\chi h^2
\gsim 0.12$. Along the red curves the local dark matter density predicted by our model saturates the
experimental value, i.e., $\Omega_{\text{model}}h^2 = 0.12$.

The solid and dashed green curves are, respectively, bounds from 
the 90\% C.L. exclusion limits set by LUX~\cite{Akerib:2013tjd} 
for spin-independent cross-sections and XENON100~\cite{Aprile:2013doa} 
for spin-dependent cross-sections assuming the canonical local 
dark matter density of $\rho_\chi \simeq 0.3$~GeV/cm$^3$.
For a purely thermal origin of the dark fermions, this bound 
only applies at the crossing of the green curves with the red curve. 
Above the red curve, the green lines correspond to the 
constraint on the parameter space assuming there is some other
origin of the dark fermion abundance that makes up the correct
cosmological density (and thus local density) that we observe today.
The shaded green region, by contrast, is ruled out even if 
the abundance of the dark fermions \emph{is} the predicted (subdominant)
thermal abundance associated with those parameters.
In this case, even if there were another (inert) component of
dark matter to make up the difference in relic density, 
the small thermal abundance of the dark fermions 
($\propto 1/\lambda^4$) is compensated by 
an enhanced direct detection scattering cross section ($\propto \lambda^4$).  

In the following we remark on the various features of the constraints in
Figs.~\ref{fig:AllConstraintsu}, \ref{fig:AllConstraintsd} and \ref{fig:AllConstraintsud}.

We first note that the dilepton spectrum constraints are generically tighter for Models U and UD
than for \modd. This follows from the PDFs of the initial state up quarks in comparison to
initial state down quarks, leading to higher production rates when the former are present in the
new physics process. Note that the constraints from the dilepton spectrum lie in the region where the new physics
signal is dominated by $|\mathcal{M}_{\text{box}}|^2$, hence the largest contributions
to the significance arise from the region around $\MLL \simeq M_1 + M_2$. For the
set of parameters spanned by the blue curve, this does not give rise to a significant
difference between the $\delM = 0.1$ and $\delM = 0.5$ cases
in all three models, as can be observed comparing the left- and right-hand sides of the figures.

We also note that the dilepton spectrum constraints are stronger when the mass splitting of the mediator and dark matter is small. This is because the monocline is sharper for a degenerate spectrum as demonstrated in Fig.~\ref{fig:pDiracXSMassesVaried}. A mass splitting between $\phi$ and $\chi$ causes a transition from an SM-like spectrum in the IR to the parallel SM+DM-like spectrum in the UV over a larger mass interval. In contrast, searches for mediator pair production in jets+MET events become weaker for smaller $M_\phi/M_\chi$ due to the reduced amount of missing transverse energy. This demonstrates the complementarity of our  dilepton spectrum observables to existing DM searches.

In all three models, one finds the jets+MET constraints slightly stronger for $\delM = 0.5$ than
$\delM = 0.1$. The reasons for this behavior are outlined in detail in~\cite{Kribs:2013eua}, but we
summarize it here as follows. Same-handed squark production, such as in the first
Feynman diagram in Fig.~\ref{fig:FeynMJETs}, is absent in the pure Dirac limit, but turns on gradually
as we approach the Majorana limit, contributing to the production rates. Hence the jets+MET bounds
tighten as we increase $\delM$. Once again due to PDF effects this search sets tighter constraints
on \modud~than \modu, which in turn are tighter than in \modd. The cuts used in the search get more
efficient when the scalar mediator and the LSP are more split in mass. This dependence
on the acceptance gives rise to the strengthening of the bounds observed as $\Mrat$ increases. For \modd, the acceptance for a near-degenerate
spectrum is poor enough to set no bounds at all in our chosen range of $\lambda$ for $\Mrat$.
In all the plots, we have assumed that the production of both $\chi_1$ and $\chi_2$ contributes
to the MET, while realistically $\chi_2$ would undergo a decay to $\chi_1$ and SM states.

The direct detection limits contain several interesting features.
First, the bounds are identical for either splitting, $\delM = 0.1, 0.5$, since for a
splitting of more than $\mathcal{O}(100 \; {\rm keV})$, there is insufficient kinetic energy in the nonrelativistic collisions to excite to the heavier state $\chi_2$.  Hence, $\chi_1$ behaves entirely Majorana-like for direct detection searches.
Next, the spin-independent (SI)
scattering bounds (solid green curves) are very similar for \modu~and D, but stronger for \modud~since
more partons are involved in the scattering in the latter. The spin-dependent (SD) bounds (dashed
green curves)
differ across all three Models due to the difference in nucleon matrix elements, 
which are large for down quarks in a neutron and small for up quarks in a neutron (see Appendix~\ref{sec:ddformulae}).
Moreover, the SI constraints weaken much more rapidly than the SD bounds as
we increase in $\Mrat$ (and disappear for $\Mrat = 2$ in the range of
our parameter space). This is due to the dominance of the twist-2 operator in SI scattering
that scales as $1/M_\phi^8$, as opposed to the $1/M_\phi^4$-dependence of SD scattering. This interplay between the dimensionality of the operator and the relative strengths of
the nucleon matrix elements determines whether the SI or the SD direct detection results
sets the stronger bounds, that in turn depends on the choice of the Model and
$\Mrat$. In \modu, the SI bounds are stronger than the SD bounds up to $\Mchi \sim 1100$ GeV
for $\Mrat = 1.1$ and up to  $\Mchi \sim 500$ GeV for $\Mrat = 1.2$; for $\Mrat \geq 1.3$, the SD bounds
are stronger. In \modd, the SI bound is stronger up to  $\Mchi \sim 600$ GeV for $\Mrat = 1.1$, and
the SD bound is uniformly stronger for $\Mrat \geq 1.2$. In \modud, the SI bounds are uniformly stronger
for $\Mrat \leq 1.3$, stronger than SD bounds up to $\Mchi \sim 400$ GeV for $\Mrat = 1.5$ and weaker
(here absent) for $\Mrat = 2$.

The relic density constraints are slightly weaker for \modud~than Models U and D since a pair of
$\chi$'s can annihilate to two different flavors of quark final states.
As $\Mrat$ is increased, the relic density bounds gradually increase in all three Models. This is due to the
weak dependence of $\langle \sigma_{\text{eff}}v_{\text{rel}} \rangle$ on $\Mrat$, as can be seen from
Appendix~\ref{sec:aeffbeff}. The annihilation of Majorana dark matter happens without
an $s$-wave component due to chirality-suppression (see \cite{Chang:2013oia}) and hence is less
efficient than Dirac dark matter annihilation. For a mixed dark matter candidate like ours, $\delM=0.1$
approximates the Dirac case and $\delM=0.5$ approximates the Majorana case during freeze-out.
This is why the thermal relic bounds on the right-hand-side of  Figs.~\ref{fig:AllConstraintsu},
\ref{fig:AllConstraintsd} and \ref{fig:AllConstraintsud} are stronger than those of the left-hand-side.

Finally, we remark on the striking complementarity of the various dark matter probes applied to our model.
While it is obvious that the relic constraints bound Models U, D and UD in the low-$\lambda$ regime,
several competing factors determine which of the other experiments -- dilepton searches, jets+MET
searches, direct detection -- set the strongest bound at higher couplings $\lambda$. In fact, depending on the Model
and choice of parameters, each of these three can give the best bounds in some parameter regime. Since the
dependence of each probe on the parameters has been explained in this section, in the following we only
briefly describe our findings, as applicable to the high-$\lambda$, low-$\Mchi$ region.

 In \modu, for both $\delM = 0.1$ and 0.5, the
tightest exclusions come from direct detection for $\Mrat \leq 1.3$, direct detection for $\Mchi \lsim 450$ GeV and
dilepton measurements for $\Mchi \gsim 450$ GeV at $\Mrat = 1.5$, and predominantly jets+MET searches
at $\Mrat = 2$. In \modd, the tightest exclusions are from direct detection for all $\Mrat$.  In \modud, direct
 direction predominantly sets the tightest limits for $\Mrat \leq 1.3$; for $\Mrat =1.5$, the best bounds are
 placed by jets+MET searches at $\Mchi \lsim 500$ GeV for $\delM = 0.1$ and at $\Mchi \lsim 600$ GeV for
 $\delM = 0.5$, and by dilepton measurements at $\Mchi \gsim 500$ GeV for $\delM = 0.1$; for $\Mrat = 2$,
 the best bounds are placed by jets+MET searches at $\Mchi \lsim 500$ GeV and dilepton measurements
  at $\Mchi \gsim 500$ GeV.

\section{Future Projections} \label{sec:future}

\begin{figure*}
\begin{center}
\includegraphics[width=7cm]{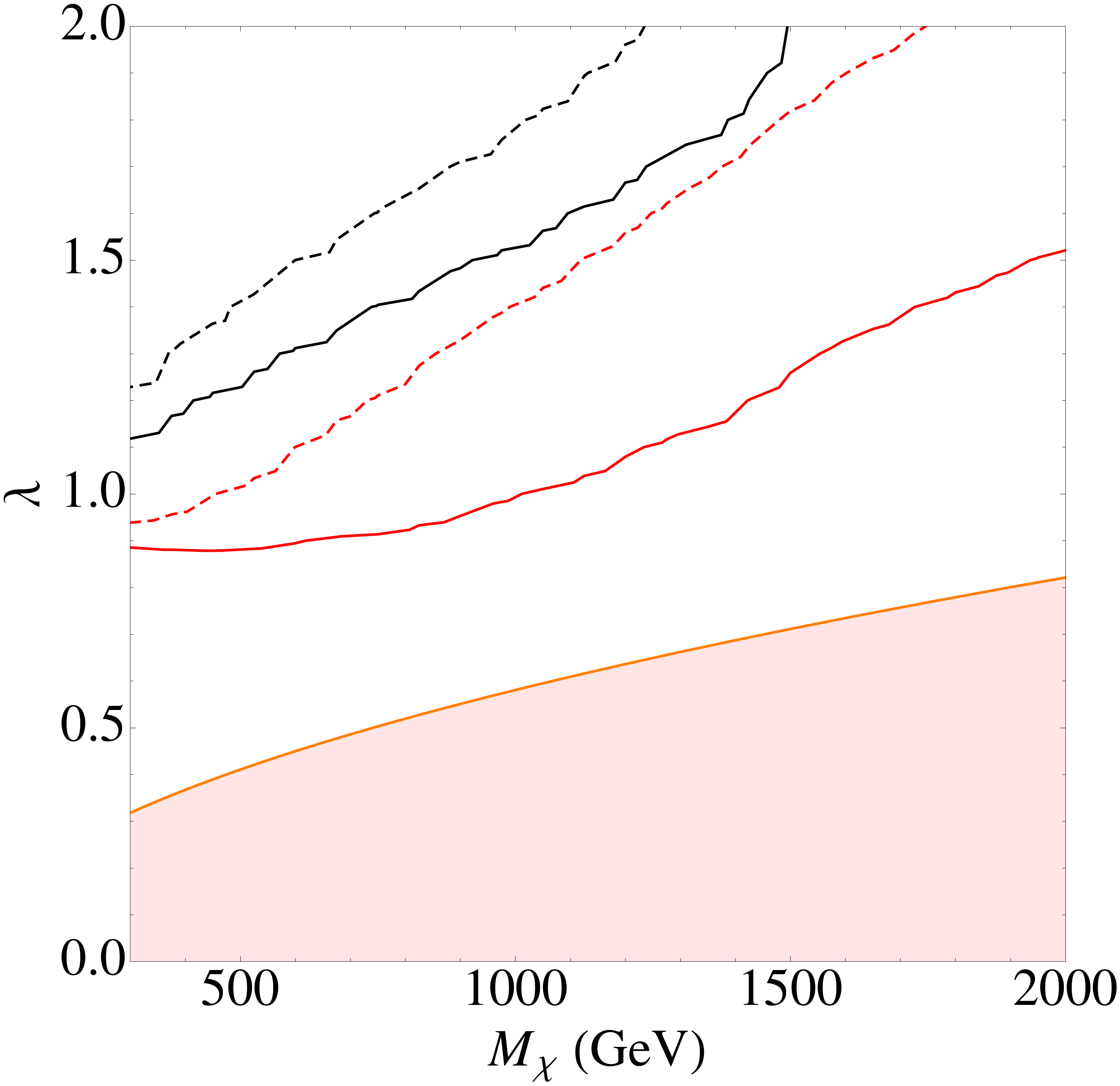}
\quad \quad \quad
\includegraphics[width=7cm]{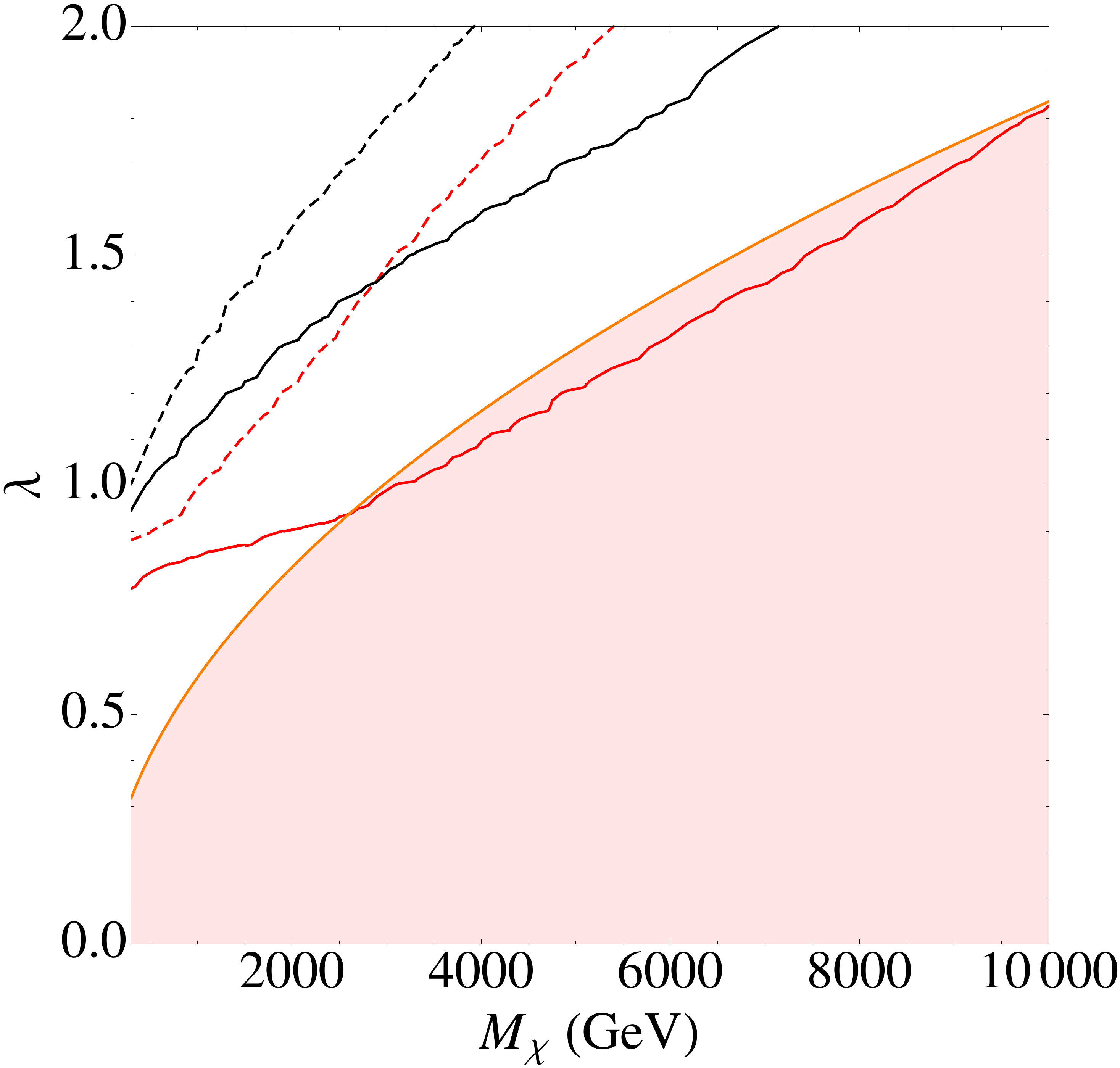}
\end{center}
\caption{Projections for 95\% C.L. sensitivity for the LHC running at $\sqrt{s}$  = 14 TeV (left) and  $\sqrt{s} $ = 100 TeV (right).
The red (black) curves denote \modu~(D). The dashed (solid) curves correspond to an integrated luminosity of $100 \text{fb}^{-1} (3000 \text{fb}^{-1})$.}
\label{fig:projexns}
\end{figure*}

We provide in Fig.~\ref{fig:projexns} our projections for the sensitivity of 
the LHC at $\sqrt{s} = 14$~TeV (left) and for a future proton-proton collider at $\sqrt{s} = 100$~TeV (right) in the dilepton invariant mass spectrum.
Here we have chosen $\Mrat = 1.2$ and $\delM = 0.1$ for \modu~and \modd, 
for illustration. We expect other choices of parameters
to not qualitatively alter the results presented, as may be deduced from the $\Delta \chi^2$ bounds shown across
various sets of parameters in Figs.~\ref{fig:AllConstraintsu}, \ref{fig:AllConstraintsd} and \ref{fig:AllConstraintsud}.   The
 shaded red region corresponds to an overabundance of dark matter ($\Omega_\chi h^2
\gsim 0.12$) for $\delM = 0.1$. The dashed (solid) curves correspond to an integrated luminosity
of $100 \; \text{fb}^{-1} (3000 \; \text{fb}^{-1})$; the red (black) curves correspond to \modu~(D).

To obtain these plots, leading order cross-sections were computed using MadGraph5 with CTEQ6L1 parton
distribution functions and a global K-factor of 1.25 was applied to obtain projected background
events. Such a procedure may not capture all the considerations that may go into computing the background
for a 100 TeV collider. For a full analysis, for instance, one would need to compute the effects of the
double logarithmic contributions of Sudakov electroweak corrections~\cite{Hook:2014rka}, take into account the (modified) running of the standard model gauge couplings~\cite{Alves:2014cda}, etc.
(For additional considerations of dark matter physics at 
100 TeV, see for example \cite{Cohen:2013xda,Low:2014cba,Gori:2014oua}.)
Our objective here is to present sensitivity projections that are indicative of what one might expect with extrapolations of what has been done already at the LHC at $\sqrt{s} = 8$~TeV\@. 
In particular, a uniform, uncorrelated systematic error of $6\%$ was assumed across all bins. ``Signal'' events were
generated by running 100 pseudo-experiments, applying Poisson fluctuations around the background events.
A $\Delta \chi^2$-fit, as defined in Eq.~(\ref{eq:DeltaChiSq}), was then performed with each pseudo-experiment's results and
the arithmetic mean of the $\Delta \chi^2$'s was obtained.
95\% C.L. exclusion limits were then set on the $\lambda-M_\chi$ plane.

As one can see, the dilepton spectrum features are significantly more
prominent at the LHC at 14 TeV, and even more so at a
100 TeV future collider. This is to be expected since the number of dilepton events increases considerably both
by the higher center-of-mass energy and the higher integrated luminosities, thus improving the sensitivity
of a shape-fit. For the same reason, the systematic uncertainties, which were smaller than statistical
uncertainties in the $\sqrt{s} = 8$ TeV
 measurements and hence negligible in setting constraints,
play a more important role in determining the sensitivities of future colliders. 

One also notices the difference in slope between the solid red curves 
and the others in the left-hand plot, and between solid and dashed curves 
in the right-hand plot. This is because the contributions to 
$\Delta \chi^2$ come from a wider range near $\MLL \simeq 2 \Mchi$.

We do not attempt to make projections for jets+MET constraints 
on the model.\footnote{We are very grateful to 
Gavin Salam and Andreas Weiler for discussions about the
applicability (and difficulties) of using their (awesome) 
Collider Reach tool~\cite{colliderreachtool} for projections in this squeezed scenario.}  
This is because the exclusion cross-sections obtained by recasting the supersymmetry searches are extremely sensitive to the 
choice of cuts in the phase space when the LSP and colored scalar 
are nearly degenerate in mass.  We anticipate some complementarity
between the jets+MET sensitivity and dilepton sensitivity:
as $\Mrat$ is increased, the dilepton monocline signal is 
suppressed, while the jets+MET signal becomes more easily visible.
Where the crossover occurs is undoubtedly highly sensitive to 
the respective detection search strategies.  

\section{Discussion and Outlook}

We have presented a simple but realistic model of pseudo-Dirac 
fermionic dark matter that results in a qualitatively new signal
in the form of kinematical and angular features in 
dilepton production at the LHC\@. 
The most spectacular feature is the ``monocline'', a step-like feature
with a sharp rise in the differential cross section for dilepton
production occurring for a dilepton invariant mass near the sum of the 
dark fermion masses, $\MLL \sim M_1 + M_2$, with a subsequent
gradual falloff.  If discovered, this signal provides an immediate 
target of opportunity given that the putative dark matter particle's 
mass is bounded 
(namely, $m_{\rm DM} \lsim \MLL/2$ for a monocline feature at $\MLL$).  
Of course observing the feature consistent with a radiative correction
from a box of new particles with masses $\sim \MLL/2$ does not 
immediately imply these particles are dark matter.
Nevertheless, knowing the scale is immensely useful when
applied to direct and indirect detection experiments, 
as well as traditional signals at colliders of both the dark matter
(e.g., mono-X $+$ MET signals) as well as the scalar mediators
(e.g., jets $+$ MET for the scalar mediator).  We also note that our monocline signal is most powerful when the spectrum of dark matter and its mediator is nearly degenerate. This strategy is thus complementary to MET-based searches.

Pseudo-Dirac dark matter, that we have shown leads to interesting 
signals in dilepton production, is also well-motivated and predictive.  
Pseudo-Dirac fermions 
could arise naturally when an accidental 
$U(1)$ symmetry that gives a Dirac mass to the fermions is broken 
at loop level~\cite{Weinberg:1972fn}. Since a Dirac fermion can be 
thought of as two degenerate Majorana eigenstates, 
the effect of the small Majorana mass is to introduce a splitting
in the eigenmasses. If the splitting is of the order of a few GeV, 
we obtain several desirable features.  Among these is that
since the momentum transfer scale of direct detection experiments 
is $10$-$100$s of keV, such experiments are only sensitive to the 
lighter eigenstate; thus, the pseudo-Dirac fermion with a 
few-GeV-splitting can be treated as a Majorana fermion for 
direct detection.  In addition, efficient $s$-wave coannihilation 
between the two eigenstates would result in a relic abundance 
that does not overclose the universe even for small couplings.
The heavier eigenstate produced in a collider can decay to the 
lighter one with a displaced vertex that is measurable at the
LHC. By studying the dilepton spectrum in this decay, 
the mass splitting can be directly measured. 
The decay length can also predict the mass of the lighter
state if the model's relic abundance is matched with the observed 
value and if the mediators are heavy. 

The model as presented is renormalizable, and thus in principle 
UV complete.  However, we have considered relatively large 
(though perturbative) $\lambda$ couplings between the dark fermion,
the scalar mediator, and a Standard Model quark or lepton.
These couplings, when RG evolved to higher scales, may develop
Landau poles.  This is not in itself a concern for us since
we have focused on the physics of the new particles near
to their threshold production at the LHC\@.  
Larger $\lambda$ couplings could arise from several sources.
The most logical possibility is that there is a larger set of
scalar mediators, for instance scalar quark mediators that
couple to the left-handed quarks, that, when summed into the 
box contributions, masquerade as a larger effective $\lambda$ 
coupling with fewer mediators.   
Another possibility is that the pseudo-Dirac fermionic 
partner $\chi_A$ couples to the scalar mediators and quarks,
also effectively increasing the strength of the $\lambda$ couplings.

The model has unmistakable similarities to simplified
supersymmetric models with a bino or neutral wino as the
dark matter, with the squarks and sleptons are the scalar 
mediators.  Indeed, the supersymmetric limit 
is interesting, since several of our otherwise arbitrary
assumptions (coupling of just $\chi_B$ to the scalar mediator 
and quarks) could arise naturally in a supersymmetric context.  
The main impediment is that an observable feature in
dilepton production requires $\lambda \gtrsim g$ by a factor
of perhaps $1.5$ -- $3$ times what would have otherwise been 
required by (at least unbroken) supersymmetry. 
This is intriguingly reminiscent of the Higgs quartic coupling,
which is related to the electroweak couplings at tree-level, 
but in fact must be significantly larger to accommodate 
the observed value of $125$~GeV\@.  An interesting
question for future exploration is to understand what 
could be possible from supersymmetry breaking corrections 
to increase the size of the quark-squark-neutral gaugino coupling.

We have not considered flavor-violation in the model, 
but this too could be interesting, especially if the 
dilepton signal was also accompanied by some fraction of 
$e^\pm \mu^\mp$ events (that would also exhibit a feature
in their $\MLL$ spectrum).  We did not consider flavor-violation
in this paper for two reasons:  one is that it obviously 
would not interfere with SM Drell-Yan production,
which was our primary motivation.  
Second, we would necessarily be forced into considering 
additional lepton-flavor-violating constraints, which are likely
to be highly constraining. For a discussion of quark flavor constraints on models similar to ours see~\cite{Agrawal:2014aoa}.

The scalar quark mediator will necessarily have box contributions
to the dijet signal as well.  Unfortunately, our estimates
of the size of this radiative correction are that it is much 
too small to lead to an observable monocline signal in the 
dijet spectrum.  This is because the box 
contribution arises in the partonic process 
$q\bar{q} \ra q\bar{q}$ whereas the dominant dijet
production involves $q q \ra q q$ as well as gluon mediated processes,
which are much more significant given the associated PDF enhancements. 
To get a signal that could compete with QCD strength would 
require $\lambda$ couplings much larger than required 
for dilepton production, and this suggests that a perturbative
analysis is no longer possible.

In summary, we encourage ATLAS and CMS to explore the sensitivity
of features in the dilepton kinematic and angular spectrum 
for extracting dark matter signals!

\begin{acknowledgments}

We are very grateful to 
Yang Bai, 
Joshua Berger,
Spencer Chang,
Adam Martin, 
Arjun Menon, 
Gilad Perez,
Gavin Salam, 
Seema Sharma,
Tim Tait, and 
Andreas Weiler
for useful discussions and comments during the course of this work.
NR thanks the 43rd SLAC Summer Institute, where
part of this work was completed, for inspiring discussions.
This work was supported in part by National Science Foundation Grant 
No. PHYS-1066293 and the hospitality of the Aspen Center for Physics. 
WA at Perimeter Institute is supported by the Government of  
Canada through Industry Canada and by the Province of Ontario through  
the Ministry of Economic Development \& Innovation.  
PJF and RH are supported by Fermilab,  
operated by Fermi Research Alliance, LLC, under  
contract DE-AC02-07CH11359 with the United States Department of Energy. 
GDK and NR are supported in part by the Department of Energy
under contract numbers DE-FG02-96ER40969 and DE-SC0011640.

\end{acknowledgments}

\appendix
\section{Parton Level Cross-Sections} \label{app:xsections}

In this appendix, we provide expressions for the new physics box contributions to the parton level $q \bar q \to \ell^+\ell^-$ cross-sections
that are then convoluted with parton distribution functions
to obtain the proton-level differential cross-sections $d\sigma/d\MLL$.

We define the following short hand notation for 4-point loop functions
\begin{eqnarray}
D_i &\equiv& D_i[m_q^2,m_q^2,m_l^2,m_l^2,s,t,\mu_1^2,M_\phi^2,\mu_2^2,M_\phi^2] ~,~~~~ \nonumber \\
\tilde{D}_i &\equiv& D_i[m_q^2,m_q^2,m_l^2,m_l^2,s,t,\mu_3^2,M_\phi^2,\mu_4^2,M_\phi^2] ~,~~~~ \nonumber \\
\bar{D}_i &\equiv& D_i[m_q^2,m_q^2,m_l^2,m_l^2,s,u,\mu_1^2,M_\phi^2,\mu_2^2,M_\phi^2] ~,~~~~ \nonumber \\
\tilde{\bar{D}}_i &\equiv& D_i[m_q^2,m_q^2,m_l^2,m_l^2,s,u,\mu_3^2,M_\phi^2,\mu_4^2,M_\phi^2] ~,~~~~
\end{eqnarray}
with the conventions for 4-point functions as in~\cite{Hahn:1998yk}.

To incorporate the mixing of the dark fermions, we define the function
\begin{equation}
\vartheta[x] \equiv 1- \frac{x}{M_1+M_2} ~,
\end{equation}
so that $\vartheta[M_1] = \cos^2\theta$ and $\vartheta[M_2] = \sin^2\theta$.
Here, $\theta$ is the mixing angle introduced in Eq.~(\ref{eq:eigstats}).
In the following, $c=2/3$ for up quarks in the initial state
and $c=-1/3$ for down quarks in the initial state.

The interference of the tree-level
$s$-channel photon-mediated diagram with

(i) any direct box diagram is given by
\begin{eqnarray}
&& d\tilde{\sigma}_{\gamma-\text{box}}[\mu_1,\mu_2] =
-\vartheta[\mu_1]\vartheta[\mu_2]\frac{ c e^2 |\lsq|^2|\lsl|^2}{256 \pi^3} \nonumber \\
&& \times 2\text{Re} \left\{\frac{(s+t)^2}{s^2}
 (2D_{00} + (D_2+ D_{12} + D_{22} + D_{23})s ) \right\} ~;
 \nonumber
\end{eqnarray}

(ii) any crossed box diagram is given by
\begin{eqnarray}
&& d\tilde{\sigma}_{\gamma-\text{xbox}}[\mu_1,\mu_2] = -\vartheta[\mu_1]\vartheta[\mu_2] \frac{ c e^2 |\lsq|^2|\lsl|^2}{256 \pi^3}
\nonumber \\
&& ~~~~~~~~~~~~~~~ \times 2\text{Re}\left\{\frac{(s+t)^2}{s^2} (\mu_1 \mu_2 \bar{D}_0)\right\}
~. \nonumber
\end{eqnarray}

The interference of the tree-level
$s$-channel $Z$-mediated diagram with

(i) any direct box diagram is given by
\begin{eqnarray}
&& d\tilde{\sigma}_{Z-\text{box}}[\mu_1,\mu_2] = -\vartheta[\mu_1]\vartheta[\mu_2] \frac{c e^2 t_W^2
|\lsq|^2|\lsl|^2}{256 \pi^3}
\nonumber \\
 && \times 2\text{Re}\left\{ \frac{(s+t)^2}{s (s-M_Z^2)} (2D_{00} + (D_2+ D_{12} + D_{22} + D_{23})s) \right\}~;
 \nonumber
 \end{eqnarray}

(ii) any crossed box diagram is given by
\begin{eqnarray}
&& d\tilde{\sigma}_{Z-\text{xbox}}[\mu_1,\mu_2] = -\vartheta[\mu_1]\vartheta[\mu_2] \frac{c e^2 t_W^2 |\lsq|^2|\lsl|^2}{256 \pi^3}
\nonumber \\
&& ~~~~~~~~~~~~~~ \times 2\text{Re}\left\{\frac{(s+t)^2}{s (s-M_Z^2)} (\mu_1 \mu_2 \bar{D}_0)\right\} ~,
\nonumber
\end{eqnarray}
where $t_W = \tan\theta_W$ and $\theta_W$ is the weak mixing angle.

Thus, the interference between the all the tree diagrams and any direct
box diagram is
\begin{eqnarray}
&& d\tilde{\sigma}_{\text{tree}-\text{box}}[\mu_1,\mu_2] = \nonumber \\
&& ~~~~~~~~ d\tilde{\sigma}_{\gamma-\text{box}}[\mu_1,\mu_2]
+d\tilde{\sigma}_{Z-\text{box}}[\mu_1,\mu_2] ~,
\label{eq:intalltreeanybox}
\end{eqnarray}
and the interference between all the tree diagrams and any crossed box
diagram is
\begin{eqnarray}
&& d\tilde{\sigma}_{\text{tree}-\text{xbox}}[\mu_1,\mu_2] = \nonumber \\
&& ~~~~~~~ d\tilde{\sigma}_{\gamma-\text{xbox}}[\mu_1,\mu_2]
+d\tilde{\sigma}_{Z-\text{xbox}}[\mu_1,\mu_2] ~.
\label{eq:intalltreeanyxbox}
\end{eqnarray}
The interference between any two direct box diagrams is given by
\begin{eqnarray}
&&d\tilde{\sigma}_{\text{box}^2}[\mu_1,\mu_2,\mu_3,\mu_4]=
\nonumber \\
&&~~~~~~~~ \vartheta[\mu_1]\vartheta[\mu_2]
\vartheta[\mu_3]\vartheta[\mu_4] \frac{|\lsq|^4|\lsl|^4}{2048\pi^5s}(s+t)^2 \nonumber \\
&&\times 2\text{Re} \Big\{ (2D_{00} + (D_2+ D_{12} + D_{22} + D_{23})s)
\nonumber \\
&&~~~~~~\times (2\tilde{D}^*_{00} +(\tilde{D}^*_2+ \tilde{D}^*_{12} + \tilde{D}^*_{22} + \tilde{D}^*_{23})s)\Big\} ~.
\label{eq:intanyboxanybox}
\end{eqnarray}
The interference between any two crossed box diagrams is given by
\begin{eqnarray}
&&d\tilde{\sigma}_{\text{xbox}^2}[\mu_1,\mu_2,\mu_3,\mu_4]=\nonumber \\
&& ~~~~~~~ \vartheta[\mu_1]\vartheta[\mu_2]\vartheta[\mu_3]\vartheta[\mu_4] \frac{|\lsq|^4|\lsl|^4}{2048\pi^5s}(s+t)^2
\nonumber \\
&& ~~~~~~~~~~~~~~~~~~~~~~ \times 2\text{Re} \Big\{ (\mu_1 \mu_2 \mu_3 \mu_4 \bar{D}_0 \tilde{\bar{D}}^*_0) \Big\} ~.
\label{eq:intanyxboxanyxbox}
\end{eqnarray}
The interference between any direct box diagram and any crossed box
diagram is given by
\begin{eqnarray}
&&d\tilde{\sigma}_{\text{box}-\text{xbox}}[\mu_1,\mu_2,\mu_3,\mu_4]= \nonumber \\
&& ~~~~~~~ \vartheta[\mu_1]\vartheta[\mu_2]\vartheta[\mu_3]\vartheta[\mu_4]\frac{|\lsq|^4|\lsl|^4}{2048\pi^5s}(s+t)^2
\nonumber \\
&& \times 2\text{Re} \Big\{
 ( (2D_{00} + (D_2+ D_{12} + D_{22} + D_{23})s) \nonumber \\
&& ~~~~~~~~~~~~~~~~~~~~~~~~~~~~~~~~~~~~~~~~~ \times \mu_3 \mu_4 \tilde{\bar{D}}^*_0) \Big\} ~.
  \label{eq:intanyboxanyxbox}
\end{eqnarray}

We can now write down the total cross-sections using the expressions above.
From Eq.~(\ref{eq:intalltreeanybox}), the interference between all the tree
diagrams and all the direct box diagrams is obtained as
\begin{equation}
\nonumber
d\sigma_{\text{tree}-\text{box}} = \sum_{a,b=1,2}d\tilde{\sigma}_{\text{tree}-\text{box}}[M_a,M_b] ~.
\end{equation}
From Eq.~(\ref{eq:intalltreeanyxbox}), the interference between
all the tree diagrams and all the crossed
box diagrams is obtained as
\begin{equation}
\nonumber
d\sigma_{\text{tree}-\text{xbox}} = \sum_{a,b=1,2}d\tilde{\sigma}_{\text{tree}-\text{xbox}}[M_a,M_b] ~.
\end{equation}
From Eq.~(\ref{eq:intanyboxanybox}), the total interference
between a pair of direct boxes (including box$^2$
pieces) is given by
\begin{eqnarray}
\nonumber d\sigma_{\text{box}^2} = \frac{1}{2} \sum_{a, b, c, d = 1,2}  d\tilde{\sigma}_{\text{box}^2}[M_a,M_b,M_c,M_d] ~.
\end{eqnarray}
From Eq.~(\ref{eq:intanyxboxanyxbox}), the total interference between a pair of crossed boxes (including crossed
box$^2$ pieces) is given by
\begin{eqnarray}
\nonumber d\sigma_{\text{xbox}^2} =
\frac{1}{2} \sum_{a, b, c, d = 1,2}  d\tilde{\sigma}_{\text{xbox}^2}[M_a,M_b,M_c,M_d] ~.
\end{eqnarray}
Finally, from Eq.~(\ref{eq:intanyboxanyxbox}), the total interference between direct and crossed boxes is given by
\begin{eqnarray}
\nonumber d\sigma_{\text{box}-\text{xbox}} =
\sum_{a, b, c, d = 1,2} d\tilde{\sigma}_{\text{box}-\text{box}}[M_a,M_b,M_c,M_d] ~.
\end{eqnarray}

\section{Calculation of \texorpdfstring{$a_{\text{eff}}$}{aeff} and \texorpdfstring{$b_{\text{eff}}$}{beff}}
\label{sec:aeffbeff}

In this appendix we describe the calculation of  $a_{\text{eff}}$ and $b_{\text{eff}}$,
which characterize the $s-$wave and $p-$wave contributions to the effective
annihilation cross-section $\langle \sigma_{\text{eff}}v_{\text{rel}} \rangle$ of pseudo-Dirac
dark matter.

Consider the annihilation process $\chi_1 \chi_2 \ra f \bar{f}$, which proceeds
through the $t$ and $u$ channels. Here $\chi_1$ and
$\chi_2$ are two Majorana fermions with masses $\mu_1$ and $\mu_2$ respectively,
 and $f$ is a SM fermion taken to be massless for simplicity.
Taylor-expanding in $v$ to write $\langle \sigma v \rangle = a[\mu_1,\mu_2]
+ b[\mu_1,\mu_2]v^2 + \mathcal{O}(v^4)$, we get
%
%
\begin{eqnarray}
a[\mu_1,\mu_2] &=& \frac{\lambda^4 d^2(|p|-p)}{16\pi|p|(|p|+M_\phi^2)^2} ~,
\nonumber \\
b[\mu_1,\mu_2] &=& \frac{\lambda^4\text{sgn}(p)}{96\pi(|\mu_1|+|\mu_2|)(|p|+M_\phi^2)^4}
 \nonumber \\
 &\times& \Big\{ 4|p|[4 p^2 q+ p (4M_\phi^4 -q^2)+
   3M_\phi^2 d^2 q] \nonumber \\
 && + p^2[4(p-q)^2+3q^2] + M_\phi^4[3q^2+8qp-12p^2] \nonumber \\
 && - 2pM_\phi^2 d^2 [5q-2p] \Big\} ~, \label{eq:ab_lepton}
\end{eqnarray}
where $d = \mu_1 - \mu_2$, $p = \mu_1\mu_2$ and $q = \mu_1^2 + \mu_2^2$.
The expressions above hold for annihilation into leptons.
For annihilation into quarks, the expressions must be multiplied by a color
factor of 3. We recover the Majorana limit by setting $\mu_1 = \mu_2 = M_\chi$
\begin{eqnarray}
a_{\text{Maj}} &=& 0 ~,\nonumber \\
b_{\text{Maj}} &=& \frac{\lambda^4 M_\chi^2(M_\phi^4+M_\chi^4)}{12\pi(M_\phi^2+M_\chi^2)^4} ~.
\end{eqnarray}

The Dirac limit
can be obtained in the limit $-\mu_1 = \mu_2 = \Mchi$.
We obtain
\begin{eqnarray}
a_{\text{Dirac}} &=& \frac{\lambda^4 M_\chi^2}{8\pi(M_\phi^2+M_\chi^2)^2} ~, \nonumber \\
b_{\text{Dirac}} &=& -\frac{\lambda^4 M_\chi^2(-M_\phi^4+3M_\phi^2 M_\chi^2+M_\chi^4)}{24\pi(M_\phi^2+M_\chi^2)^4} ~,
\end{eqnarray}
in agreement with~\cite{Chang:2013oia} up to a factor of 4 coming from an extra $\sqrt{2}$ in the
definition of our coupling in the Lagrangian.

Let us now compute $a_{\text{eff}}$ and $b_{\text{eff}}$ by including the effect
of coannihilations between the two eigenstates of pseudo-Dirac dark matter. We do
this by making an appropriate replacement of the coupling in $a[\mu_1,\mu_2]$ and
$b[\mu_1,\mu_2]$ to account for the mixing, multiplying each term by the
appropriate Boltzmann factor and finding the weighted average.
Therefore, from Eq.~(\ref{eq:effannihilationXS}),
we have
\begin{eqnarray}
a_{\text{eff}}(x) &=&  [c_\theta^4 a[M_1, M_1]+2c_\theta^2s_\theta^2 a[M_1, M_2]w(x)
\nonumber \\
&& +s_\theta^4 a[M_2, M_2]w^2(x)]/[(1+w(x))^2] ~,
\nonumber \\
b_{\text{eff}}(x) &=&  [c_\theta^4 b[M_1, M_1]+2c_\theta^2s_\theta^2 b[M_1, M_2]w(x)
\nonumber \\
&& +s_\theta^4 b[M_2, M_2]w^2(x)]/[(1+w(x))^2] ~,
\end{eqnarray}
where $w(x) = (1+\delta)^{3/2}e^{-x \delta}, \delta = (M_2 - M_1)/M_1$.
\\

\section{Direct Detection Formulae}
\label{sec:ddformulae}

We follow the approach of \cite{Hisano:2010yh} (see also \cite{Drees:1993bu})
to compute the spin-independent scattering cross-section
of $\chi_1$ with nucleons, obtained as

\begin{equation}
\sigma_{\text{SI}} = \frac{4}{\pi} \mu_N |f_N|^2
\end{equation}
where $\mu_N$ is the $\chi-N$ reduced mass (N=p,n) and $f_N$ is given by
\begin{eqnarray}
\nonumber \frac{f_N}{m_N} &=&  \sum_{q=u,d} \left( f_q f_{T_q}
 + \frac{3}{4} (q(2) + \bar{q}(2)) g_q \right) ~,
\end{eqnarray}
with $f_q = \lambda^2 \Mchi/[16(\Msq^2 - \Mchi^2)^2]$, $g_q = 4 f_q$.
Only the quarks that couple to our dark sector are included in the summations given here. The nucleon
matrix elements of the quark operators are taken from \cite{Chang:2013oia} (see also~\cite{Crivellin:2013ipa}).
The large values of $q(2)+ \bar{q}(2)$ make the quark twist-2 contribution the dominant one.

The spin-dependent cross-section for scattering between nucleons and $\chi$ is given by \cite{An:2013xka}
\begin{equation}
\sigma^{\text{Maj}}_{\text{SD}} = \frac{3}{64\pi} \frac{\lambda^4\mu^2_N(\sum_q \Delta_q^N)^2}{(M^2_{\tilde{q}}-\Mchi^2)^2} ~,
\label{eq:sigmasd}
\end{equation}
where the summation is again over the quarks that couple to the hidden sector and $\Delta^N_q$ is defined by
$2s_\mu \Delta^N_q \equiv \langle N| \bar{q} \gamma_\mu \gamma_5 q|N \rangle $ with $s_\mu$ the nucleon
spin operator.  We take bounds from the neutron-dark matter scattering
since they are stronger, thus the appropriate matrix elements we use are $\Delta_u^n = -0.427, \Delta_d^n = 0.842.$ \cite{Airapetian:2006vy}.


\end{document}